\documentclass[aps,prd,twocolumn,showpacs,unsort]{revtex4-1}

\usepackage[top=1in, bottom=1in, left=1in, right=1in]{geometry}

\usepackage{amsmath}
\usepackage{amsfonts}
\usepackage{wasysym}
\usepackage{graphicx}
\usepackage{comment}
\usepackage{tensor}

\def\filetype{pdf}

\allowdisplaybreaks[1]

\begin{document}

\title{Stability and Critical Behavior of Gravitational Monopoles}
\author{Ben Kain}
\affiliation{Department of Physics, College of the Holy Cross, Worcester, Massachusetts 01610, USA}

\begin{abstract}
\noindent I dynamically evolve spherically symmetric spacetimes containing gravitational 't Hooft--Polyakov monopoles and determine the stable end states of the evolutions.  I do so to study stability and critical behavior of the well-known static gravitational monopole solutions.  For the static solutions, there exist regions of parameter space where two static monopole black holes and the static Reissner-N\"ordstrom black hole have the same mass.  I find strong evidence that one of the static monopole black hole solutions is a critical solution, to which near-critical solutions are dynamically attracted before evolving to one of the other two static solutions as end states.  I also discuss the no-hair conjecture for this model in the context of collapse.
\end{abstract} 

\maketitle


\section{Introduction}

Magnetic monopole solutions can be found in some spontaneously broken non-Abelian gauge theories \cite{Goddard:1977da}.  First discovered in its simplest adaptation, $SU(2)$ with a real triplet scalar field, the 't Hooft--Polyakov monopole is a classical solution to the equations of motion with finite energy \cite{tHooft:1974kcl, Polyakov:1974ek}.  It was subsequently generalized to curved space \cite{VanNieuwenhuizen:1975tc} and static solutions for both regular and black hole gravitational monopoles were found \cite{Lee:1991vy, Ortiz:1991eu, Breitenlohner:1991aa, Breitenlohner:1994di, Volkov:1998cc}.

Stability of the static gravitational monopole solutions is nontrivial \cite{Lee:1991qs, Aichelburg:1992st, Maeda:1993ap, Tachizawa:1994wn, Hollmann:1994fm}.  There exist regions of parameter space where two static black hole monopole solutions and the static Reissner-Nordstr\"om (RN) black hole solution all exist with the same mass and it is not so simple as the least massive solution is stable with the others unstable.  I study stability by dynamically evolving the system to determine the stable end state of the evolution.  I find that one of the static monopole black hole solutions and the static RN solution are both stable and the other static monopole black hole solution is unstable.

Black holes are well known to exhibit critical phenomena \cite{Choptuik:1992jv, Gundlach:2002sx, Gundlach:2007gc}.  It is here that unstable solutions can be particularly interesting as critical solutions, acting as intermediate attractors between two different end states.  There has been substantial study of critical solutions at the threshold of collapse \cite{Gundlach:2002sx, Gundlach:2007gc}, but comparatively less for critical solutions sitting between different black hole end states \cite{Choptuik:1999gh, Millward:2002pk, Rinne}.  I find strong evidence that the unstable black hole monopole solutions are critical solutions sitting between a stable black hole monopole and the RN black hole.

As far as I am aware, numerical simulations of the gravitational 't Hooft--Polyakov monopole system has been presented only once before in \cite{Sakai:1995ds} (for simulations in flat space see \cite{Fodor1, Fodor2}), which focused on vacuum values of the scalar field larger than or near its maximum value, above which static solutions no longer exist.  Here my interest is precisely with the static solutions and thus for smaller values of the scalar field vacuum value.  I solve for the dynamic solutions with a code making use of black hole excision techniques.  Black hole excision allows the code to be run indefinitely, even in the presence of a black hole, so that stable end states may be determined.

In the next section I present the fully time-dependent equations that contain the gravitational 't Hooft--Polyakov monopole and review boundary conditions.  In Sec.\ \ref{sec:static solutions} I review regular and black hole static solutions of gravitational monopoles including their stability.  In Sec.\ \ref{sec:dynamic solutions} I present dynamic solutions of gravitational monopoles, describe the code used to find them, and  study stability and critical behavior.  I also comment on the no-hair conjecture in the context of collapse.  In Sec.\ \ref{sec:conclusion} I conclude by discussing expectations for areas of parameter space not considered here.


\section{Equations and Boundary Conditions}

\subsection{Metric Equations}

Since the (flat space) 't Hooft--Polyakov monopole follows from a spherically symmetric ansatz it seems natural and convenient to restrict my study of gravitational monopoles to those in spherically symmetric spacetimes.  The general spherically symmetric metric in the ADM formalism \cite{AlcubierreBook, BaumgarteBook} is
\begin{align} \label{spherical metric}
ds^2 
&= - \left(\alpha^2 - a^2 \beta^2 \right)dt^2 + 2a^2 \beta dr dt + a^2 dr^2
\notag \\
&\qquad + B r^2 \left( d\theta^2 + \sin^2\theta d\phi^2 \right),
\end{align}
where the metric functions $\alpha$, $\beta$, $a$, and $B$ are functions of $t$ and $r$ only and I use units such that $c=1$ throughout.  $\alpha$ is the lapse and measures how quickly time moves froward from one slice to the next and $\beta$ is the only nonvanishing component of the shift vector $\beta^i=(\beta,0,0)$ in spherically symmetric spacetimes and measures how coordinates relabel themselves from once slice to the next.  The lapse and shift are gauge functions parameterizing the coordinate freedom of general relativity.  

The geometry of each slice is described by the spatial three-metric, which in spherical symmetry is diagonal: $\gamma_{ij} = \text{diag}(a^2, Br^2, Br^2\sin^2\theta)$.  The extrinsic curvature, $K\indices{^i_j}$, describes how each slice resides in the full spacetime and in spherical symmetry has only two nontrivial components, $K\indices{^r_r}$ and $K\indices{^\theta_\theta} = K\indices{^\phi_\phi}$, with the rest vanishing.  All geometric quantities obey the Einstein field equations,
\begin{equation}
G_{\mu\nu} = 8\pi G T_{\mu\nu},
\end{equation}
where $T_{\mu\nu}$ is the energy-momentum tensor to be given in the next subsection.  

Static monopole solutions are most commonly studied in radial-polar gauge.  In radial gauge the radial coordinate $r$ is chosen such that spheres of radius $r$ have area $4\pi r^2$ and is the radial coordinate used in Schwarzschild coordinates.  Radial gauge fixes $B = 1$.  Polar slicing is defined by $K\indices{^\theta_\theta} = 0$, which in conjunction with radial gauge conveniently fixes $\beta = 0$.  In radial-polar gauge the metric is described entirely in terms of the metric function $a$ and the lapse $\alpha$ which obey
\begin{equation} \label{radial-polar metric equations}
\begin{split}
\frac{a'}{a} &= 
4\pi G r a^2\rho 
- \frac{a^2 - 1}{2r}
\\
\frac{\alpha'}{\alpha} &=
4\pi G r  a^2 S\indices{^r_r} +\frac{a^2-1}{2r},
\end{split}
\end{equation}
where a prime denotes an $r$-partial derivative, which follow from the Einstein field equations.  The energy density $\rho$ and $S\indices{^r_r}$ are matter functions derived from the energy-momentum tensor and are given in the next subsection.  I use radial-polar gauge in my review of static monopole solutions in Sec.\ \ref{sec:static solutions}.

In my study of dynamic monopole solutions in Sec.\ \ref{sec:dynamic solutions} I use radial-maximal gauge.  Any numerical study in which black holes are present must take special care to avoid the singularity since infinities are disastrous for computer code.  Many techniques have been developed to avoid singularities.  I will use black hole excision methods \cite{Thornburg, Seidel:1992vd} which have the advantage that code can be run indefinitely, even in the presence of a black hole, and thus can be used to determine the stable end state of a system.  Physically anything outside a black hole is causally disconnected from anything inside the black hole and thus after a black hole forms if one removes or excises the interior region of the black hole from the simulation, and thus removes the singularity, the determination of the exterior region should be unaffected.  This presupposes that the horizon can be determined, which in fact is impossible without knowing the complete spacetime.  The standard approach is to instead excise the region inside the apparent horizon.  It is well known that coordinates in radial-polar gauge do not penetrate apparent horizons and for this reason I use radial-maximal gauge.

Radial-maximal gauge also uses radial gauge and fixes $B = 1$.  Maximal slicing is defined by $K = 0$, where $K = K\indices{^r_r} + 2 K\indices{^\theta_\theta}$ is the trace of the extrinsic curvature, and retains the shift $\beta$.  The Einstein field equations give the following equations for these three metric functions:
\begin{equation} \label{radial-maximal metric equations}
\begin{split}
a' &= 
\frac{3}{8} ra^3 (K\indices{^r_r})^2
+ 4 \pi G r a^3 \rho 
- \frac{a(a^2 - 1)}{2r} 
\\
\alpha''
&= \alpha'
\left[
 \frac{3}{8} ra^2 (K\indices{^r_r})^2
+ 4 \pi G r a^2 \rho
- \frac{a^2 +3}{2r}  
\right]
\\
&\qquad + \alpha a^2 \left[ \frac{3}{2} \left(K\indices{^r_r}\right)^2 + 4\pi G (\rho + S) \right] 
\\
K\indices{^r_r^\prime} &=  8\pi G  j_r - \frac{3}{r}K\indices{^r_r},
\end{split}
\end{equation}
where $\rho$, $S$, and $j_r$ are derived from the energy-momentum tensor and will be given in the next subsection.  I opted to write these equations in terms of $K\indices{^r_r}$ instead of $\beta$, as they are related algebraically via
\begin{equation} \label{beta Krr}
\beta = - \frac{1}{2} \alpha r K\indices{^r_r}.
\end{equation}

Regions of spacetime inside or on the boundary of an apparent horizon satisfy \cite{AlcubierreBook, BaumgarteBook}
\begin{equation}
\frac{1}{2}a r K\indices{^r_r} \leq -1.
\end{equation}
Following \cite{Choptuik:1999gh} to ensure the inner boundary lies strictly inside the apparent horizon I excise all grid points that satisfy
\begin{equation} \label{apparent horizon condition}
\frac{1}{2}a r K\indices{^r_r} \leq - \mu_H
\end{equation}
and use $\mu_H = 1.1$.

If a black hole forms and is excised the inner boundary of the computational domain moves from the origin at $r=0$ to the apparent horizon.  Inner boundary conditions for $r= 0$, which are described in Sec.\ \ref{sec:boundary conditions}, can no longer be used and a new method for obtaining boundary values is needed.  I will use the evolution equations for $a$ and $K\indices{^r_r}$, which as usual follow from the Einstein field equations, to determine their boundary values:
\begin{align} \label{a Krr evolution equations}
\dot{a} &=
- \frac{1}{2} \alpha r K\indices{^r_r} 
\biggl[
4\pi G r a^3 \rho
- \frac{a(a^2-1)}{2r}
\notag \\
&\qquad + \frac{3}{8} r a^3 (K\indices{^r_r})^2 \biggr]
 - \frac{1}{2} r a K\indices{^r_r} \alpha'
- 4 \pi G r \alpha a j_r 
\notag \\
\dot{K}\indices{^r_r} &=
\frac{3}{4} \alpha (K\indices{^r_r})^2
 - 8\pi G \alpha
 \left( S\indices{^r_r} + \frac{1}{2} r K\indices{^r_r} j_r \right) 
\notag \\
&\qquad  - \frac{\alpha(a^2- 1)}{r^2 a^2}
 + \frac{2}{ra^2} \alpha',
\end{align}
where a dot denotes a $t$-partial derivative.  These evolution equations could be used in general, but the constraint equations in (\ref{radial-maximal metric equations}), being ODEs instead of PDEs, are easier to solve and give more stable code.  The above evolution equations are then available for testing the consistency of results, which will be done in Sec.\ \ref{sec:BH excision}.  Unfortunately, an evolution equation for $\alpha$ does not exist and its boundary value must be determined in another way.  Again following \cite{Choptuik:1999gh} I ``freeze" the values of $\alpha$ and $\alpha'$ at the inner boundary after black hole formation.


\subsection{Matter Equations}

The matter content of the 't Hooft--Polyakov monopole is an $SU(2)$ Yang-Mills theory with a real triplet scalar field in the adjoint representation.  This introduces the gauge field $A_\mu^a$ and scalar field $\phi^a$, where $a = 1,2,3$ is the gauge index (which can equivalently be placed up or down).  For $SU(2)$ the generators satisfy $[T_a,T_b] = i\epsilon_{abc} T_c$, where $\epsilon_{abc}$ is the completely antisymmetric symbol with $\epsilon_{123} = 1$.  In the adjoint representation I define the components of the generator matrices as $\left(T_a\right)_{bc} = -i\epsilon_{abc}$ with normalization $\text{Tr}(T_a T_b) = 2\delta_{ab}$, where $\text{Tr}$ here and below indicates a trace over generator matrices.  It is common to refer to this as a Yang-Mills-Higgs theory.  Defining
\begin{equation}
\phi \equiv T^a \phi^a, \quad
A_\mu \equiv T^a A_\mu^a, \quad
F_{\mu\nu} \equiv T^a F^a_{\mu\nu},
\end{equation}
where a sum over repeated gauge indices is implied, $F^a_{\mu\nu}$ is the field strength, and
where such a definition for $\phi^a$ is possible because I am in the adjoint representation, the Yang-Mills-Higgs matter Lagrangian is
\begin{equation} \label{YMH Lagrangian}
\mathcal{L}_{YMH} = -\frac{1}{2} \text{Tr} 
\left[ \left(D_\mu \phi\right) \left(D^\mu \phi\right)
\right] - V - \frac{1}{8g^2} \text{Tr} \left( F_{\mu\nu} F^{\mu\nu} \right)
\end{equation}
where $g$ is the gauge coupling constant,
\begin{equation} \label{less standard covariant D}
\begin{split}
D_\mu \phi &= \nabla_\mu \phi - i \left[A_\mu, \phi\right]
\\
F_{\mu\nu} &= \nabla_\mu A_\nu - \nabla_\nu A_\mu - i \left[ A_\mu, A_\nu \right],
\end{split}
\end{equation}
and $V$ is the scalar potential whose form I give below.  Gauge transformations are defined by
\begin{equation}
\begin{split}
\phi &\rightarrow \phi' = U\phi U^{-1} 
\\
A_\mu &\rightarrow A_\mu' = U A_\mu U^{-1} - i\left(\nabla_\mu U \right) U^{-1}
\\
F_{\mu\nu} &\rightarrow F_{\mu\nu}' = U F_{\mu\nu} U^{-1},
\end{split}
\end{equation}
where $U = e^{-i\Lambda}$ and $\Lambda = \Lambda^a T^a$ with $\Lambda^a$ the gauge functions.

Spherical symmetry constrains the fields.  The general spherically symmetric $SU(2)$ gauge field takes the form \cite{Witten:1976ck, Bartnik:1988am, Volkov:1998cc}
\begin{equation} \label{gauge field}
\begin{split}
A_t &= T^3 u_t
\\
A_r &= T^3 u_r
\\
A_\theta &= T^1 w_1 + T^2 w_2
\\
A_\phi &= \left(-T^1 w_1 + T^2 w_2 + T^3 \cot\theta \right)\sin\theta,
\end{split}
\end{equation}
where $u_t$, $u_r$, $w_1$, and $w_2$ parametrize the gauge field and are functions of $t$ and $r$ only, and the real triplet scalar field takes the form
\begin{equation}
\phi = \frac{\varphi}{\sqrt{2}} T^3,
\end{equation}
where $\varphi$ is a canonically normalized real scalar field and is a function of $t$ and $r$ only.  There are a couple gauge equivalent ways these fields are commonly written in the literature \cite{Volkov:1998cc}.  I have chosen to write them in a gauge such that the generators $T^a$ are constant. I shall adhere to this gauge throughout.  The components of the spherically symmetric field strength are
\begin{align} \label{field strength comps}
F_{tr} &= T^3
 \left(\dot{u}_r - u'_t \right)
\notag \\
 F_{t\theta} &=
T^1 (\dot{w}_2 - u_t w_1) + T^2\left(\dot{w}_1 + u_t w_2 \right) 
\notag \\
F_{t\phi} &=
\bigl[T^2 (\dot{w}_2 - u_t w_1) - T^1\left(\dot{w}_1 + u_t w_2 \right) \bigr] \sin\theta 
\notag \\
F_{r\theta} &= 
T^1 \left(w'_2 - u_r w_1 \right)
+ T^2 \left(w'_1 + u_r w_2 \right) 
\notag \\
F_{r\phi} &=
\bigl[ T^2 \left(w'_2 - u_r w_1 \right) - T^1 \left(w'_1 + u_r w_2 \right) \big] \sin\theta 
\notag \\
F_{\theta\phi} &= 
-T^3 \left(1 - w_1^2 - w_2^2 \right)  \sin\theta.
\end{align}
The gauge fields obey a $U(1)$ invariance:
\begin{equation} \label{U(1) symmetry}
u_i \rightarrow u'_i = u_i - \nabla_i \tau, \qquad
w \rightarrow w' = w e^{-i\tau},
\end{equation}
where $i=t,r$, $w = w_1 + i w_2$, and $\tau$ is the gauge parameter.  Thus $u_t$ and $u_r$ transform as two components of a two-dimensional Abelian vector and $w$ transforms as a complex scalar.  This invariance will be made use of when I fix the gauge below.

The scalar potential for the 't Hooft--Polyakov monopole is
\begin{equation}
V = \frac{\lambda}{4} \left( \varphi^2 - v^2 \right)^2,
\end{equation}
where $\lambda$ is a constant and $v$ is the vacuum value of $\varphi$.  This scalar potential spontaneously breaks the $SU(2)$ symmetry down to $U(1)$ giving rise to massive vector bosons and a massive scalar field with masses
\begin{equation} \label{vector scalar mass}
m_V = gv, \qquad m_S = \sqrt{2\lambda} \, v.
\end{equation}

There are a number of ways to derive the matter equations of motion.  For example, they can be obtained from conservation of the energy-momentum tensor or by coupling the Lagrangian to gravity by constructing the Einsten-Yang-Mills-Higgs Lagrangian $\mathcal{L}_\text{EYMH} = \sqrt{-g} \mathcal{L}_\text{YMH}$, where $\sqrt{-g} = \alpha a B r^2 \sin\theta$ is the determinant of the metric, and then deriving the Euler-Lagrange equations.  For numerical purposes it is important to have the equations of motion in first order form.  I thus define
\begin{align} \label{1st order var def}
\Phi &\equiv \varphi'&
\Pi &\equiv  
\frac{a B}{\alpha} \left(\dot{\varphi} - \beta\Phi \right)
\notag \\
Q_1 &\equiv w_1' + u_r w_2&
P_1 &\equiv   \frac{a}{\alpha} 
  \Bigl( \dot{w}_1 + u_t w_2 - \beta Q_1 \Bigr)
\notag \\
Q_2 &\equiv w_2' - u_r w_1&
P_2 &\equiv \frac{a}{\alpha } 
\Bigl( \dot{w}_2 - u_t w_1 - \beta Q_2 \Bigr)
\notag \\
Y &\equiv \frac{B r^2}{2 \alpha a} \left(\dot{u}_r - u'_t \right).
\end{align}
I shall list the equations of motion grouped in families.  First $\varphi$, $\Phi$, and $\Pi$:
\begin{equation}
\begin{split}
\dot{\varphi} &= \frac{\alpha}{aB} \Pi + \beta \Phi \\
\dot{\Phi} &=
\partial_r \left( \frac{\alpha}{a B} \Pi + \beta \Phi \right) \\
\dot{\Pi} &= 
\frac{1}{r^2} \partial_r \left( \frac{\alpha B r^2}{a} \Phi + r^2 \beta \Pi \right)
- \alpha a B  
\frac{\partial V}{\partial \varphi}
\\
&\qquad - \frac{2\alpha a}{r^2} (w_1^2 + w_2^2) \varphi,
\end{split}
\end{equation}
then $w_1$, $Q_1$, and $P_1$:
\begin{align}
\dot{w}_1 &= \frac{\alpha}{a} P_1 - u_t w_2 + \beta Q_1 
\notag \\
\dot{Q}_1 &=
\partial_r  \left(\frac{\alpha}{a}P_1 + \beta Q_1 \right)
- u_t Q_2
 + u_r \left(\frac{\alpha}{a}P_2 + \beta Q_2 \right) 
\notag \\
&\qquad
+ w_2\frac{2\alpha a}{Br^2} Y
\notag \\
\dot{P}_1 &= \partial_r \left( \frac{\alpha}{a} Q_1 + \beta P_1 \right) - P_2 (u_t - \beta u_r )
 + \frac{\alpha}{a} u_r Q_2
\notag \\
&\qquad
 + \frac{\alpha a}{B r^2}w_1 (1 -w_1^2 - w_2^2)
- g^2 \alpha a w_1 \varphi^2,
\end{align}
and $w_2$, $Q_2$, and $P_2$:
\begin{align}
\dot{w}_2 &= \frac{\alpha}{a} P_2 + u_t w_1 + \beta Q_2
\notag \\
\dot{Q}_2 &=
\partial_r  \left(\frac{\alpha}{a}P_2 + \beta Q_2 \right)
+ u_t Q_1
- u_r \left(\frac{\alpha}{a} P_1 + \beta Q_1 \right)
\notag \\
&\qquad
- w_1 \frac{2\alpha a}{Br^2} Y 
\notag \\
\dot{P}_2 &= \partial_r \left( \frac{\alpha}{a} Q_2 + \beta P_2 \right) + P_1 (u_t - \beta u_r ) - \frac{\alpha}{a} u_r Q_1 
\notag \\
&\qquad
+ \frac{\alpha a}{B r^2}w_2 (1 -w_1^2 - w_2^2)
- g^2 \alpha a w_2 \varphi^2,
\end{align}
and finally
\begin{align}
\dot{u}_r &= \frac{2\alpha a}{Br^2}Y + u'_t 
\notag \\
\dot{Y} &= \frac{\alpha}{a} \left(w_1 Q_2 - w_2 Q_1\right) + \beta \left(w_1 P_2 - w_2 P_1 \right)
\notag \\
Y' &= w_1 P_2  - w_2 P_1.
\end{align}
Note that I do not have an evolution equation for $u_t$, which I will handle when fixing the $SU(2)$ gauge.

For the Yang-Mills-Higgs Lagrangian (\ref{YMH Lagrangian}) the energy-momentum tensor is
\begin{widetext}
\begin{equation}
T_{\mu\nu} = 
\text{Tr} 
\left[ \left(D_\mu \Phi\right) \left(D_\nu \Phi\right)\right]
-\frac{g_{\mu\nu}}{2} 
\text{Tr} 
\left[ \left(D_\sigma \Phi\right) \left(D^\sigma \Phi\right)\right] 
- g_{\mu\nu}V 
+ \frac{1}{g^2}g^{\sigma\lambda} F^a_{\mu\sigma} F^a_{\nu\lambda}
-\frac{g_{\mu\nu}}{4g^2}  F_{\sigma\lambda}^a F^{\sigma\lambda}_a.
\end{equation}
The nonvanishing components work out to be
\begin{align}
T_{tt} &=
\left( \frac{\alpha\Pi}{a B} + \beta \Phi \right)^2 
+
\frac{4 \alpha^2}{g^2B^2 r^4} \left(1 - \frac{a^2 \beta^2 }{\alpha^2} \right) Y^2  + \frac{2 \alpha^2}{g^2 a^2 B r^2} \left[ \left( P_1 + \frac{a \beta}{\alpha} Q_1 \right)^2 +
\left( P_2 + \frac{a \beta}{\alpha} Q_2 \right)^2\right]
\notag \\
&\qquad
-\left(\alpha^2 - a^2\beta^2\right)
\mathcal{L}_{YMH} 
\notag \\
T_{tr} &=
\left( \frac{\alpha\Pi}{a B} + \beta \Phi \right) \Phi  - \frac{4 a^2 \beta}{g^2 B^2 r^4} Y^2  
+ \frac{2\alpha}{g^2 a B r^2} \left[
\left(P_1 + \frac{a \beta}{\alpha} Q_1 \right) Q_1
+ \left(P_2 + \frac{ a \beta}{\alpha} Q_2 \right) Q_2
 \right]  
+ a^2 \beta \mathcal{L}_{YMH} 
\notag \\
T_{rr} &=
\Phi^2
 -\frac{ 4a^2 Y^2}{g^2B^2 r^4} + \frac{2(Q_1^2 + Q_2^2)}{g^2Br^2}
+ a^2\mathcal{L}_{YMH} 
\notag \\
T_{\theta\theta} &= 
\frac{T_{\phi\phi}}{\sin^2\theta}
=
(w_1^2 + w_2^2)\varphi^2
+ \frac{Q_1^2 + Q_2^2 - P_1^2 - P_2^2}{g^2 a^2} 
+ \frac{(1-w_1^2 - w_2^2)^2}{g^2Br^2}
+ Br^2 \mathcal{L}_{YMH} .
\end{align}
From these follow the matter functions, which in spherical symmetry are given by
\begin{equation}
\rho = n^\mu n^\nu T_{\mu\nu}, \qquad 
S\indices{^r_r} =\gamma^{rr} T_{rr}, \qquad
S\indices{^\theta_\theta} = S\indices{^\phi_\phi} = \gamma^{\theta\theta} T_{\theta\theta},
\qquad
j_r = -n^\mu T_{\mu r},
\end{equation}
along with $S = S\indices{^r_r} + S\indices{^\theta_\theta} + S\indices{^\phi_\phi}$, where $n^\mu = (\alpha^{-1}, -\alpha^{-1}\beta,0,0)$ is the timelike unit vector normal to the spatial slices, which are used in the metric equations in the previous subsection.  I find
\begin{equation}
\begin{split}
\rho &=
\frac{1}{2a^2} \left( \Phi^2 + \frac{\Pi^2}{B^2} \right)
+ \frac{(w_1^2 + w_2^2)\varphi^2}{Br^2} + V
+ \frac{(1 - w_1^2 - w_2^2)^2 }{2g^2 B^2 r^4}
+ \frac{Q_1^2 + Q_2^2 + P_1^2 + P_2^2 }{g^2a^2 Br^2}
+ \frac{2Y^2}{g^2 B^2 r^4} \\
S\indices{^r_r} &=
\frac{1}{2a^2} \left( \Phi^2 + \frac{\Pi^2}{B^2} \right)
- \frac{(w_1^2 + w_2^2)\varphi^2}{Br^2} - V
- \frac{(1 - w_1^2 - w_2^2)^2 }{2g^2 B^2 r^4}
+ \frac{Q_1^2 + Q_2^2 + P_1^2 + P_2^2 }{g^2a^2 Br^2}
- \frac{2Y^2}{g^2 B^2 r^4} 
\\
S\indices{^\theta_\theta} 
&= S\indices{^\phi_\phi}   
= \frac{1}{2a^2} \left(  \frac{\Pi^2}{B^2} - \Phi^2 \right)
-V
+ 
 \frac{(1 - w_1^2 - w_2^2)^2 }{2g^2 B^2 r^4}
 + 
  \frac{2Y^2}{g^2 B^2 r^4} \\
j_r
&= -\frac{\Phi \Pi }{a B}
- \frac{2 (Q_1 P_1 + Q_2 P_2)}{g^2 a B r^2}.
\end{split}
\end{equation}
\end{widetext}

The equations above are clearly complicated.  They can be significantly simplified as follows \cite{Choptuik:1999gh}.  First, using the residual $U(1)$ symmetry (\ref{U(1) symmetry}) I can set $u_t = 0$, which is welcome since I do not have an evolution equation for $u_t$.  The 't Hooft--Polyakov ansatz for the monopole sets the electric field to zero.  I can do the analogous thing here and make what is called the ``magnetic ansatz," which sets the electric field of the residual $U(1)$ to zero and leads to a dramatic simplification.  I note that the magnetic ansatz is not a gauge choice, but is a physical restriction of the theory.  I thus set $Y = 0$, since as can be seen in (\ref{1st order var def}) $Y$ is proportional to the $U(1)$ field strength.  Since $u_t=0$, setting $Y = 0$ amounts to $u_r$ being time-independent.  I can thus make a $U(1)$ transformation with a time-independent gauge parameter to remove $u_r$ without affecting $u_t$.  There is still some $U(1)$ symmetry that remains unfixed.  It can be used to remove either the real or imaginary part of $w = w_1 + iw_2$.  It is not difficult to show that the $t$- and $r$-derivatives of the gauge parameter that does this is proportional to $\dot{Y}$ and $Y'$, respectively, which are zero by the magnetic ansatz.  I choose to remove $w_2$.  Thus, by making the magnetic ansatz and judicious gauge choices the only nonzero matter fields are $w_1$ and $\varphi$.

As mentioned in the previous subsection I will be using radial-polar and radial-maximal spacetime gauges, both of which set the metric function $B=1$.  Setting $B = 1$ and simplifying the notation by defining $w\equiv w_1$, $Q \equiv  Q_1$, and $P \equiv  P_1$ the matter evolution equations reduce to
\begin{align} \label{reduced evolution equations}
\dot{\varphi} &= \frac{\alpha}{a} \Pi + \beta \Phi 
\notag \\
\dot{\Phi} &=
\partial_r \left( \frac{\alpha}{a } \Pi + \beta \Phi \right) 
\notag \\
\dot{\Pi} &= 
\frac{1}{r^2} \partial_r \left( \frac{\alpha  r^2}{a} \Phi + r^2 \beta \Pi \right)
- \alpha a 
\frac{\partial V}{\partial \varphi}
- \frac{2\alpha a}{r^2} w^2 \varphi 
\notag \\
\dot{w} &= \frac{\alpha}{a} P + \beta Q 
\notag \\
\dot{Q} &=
\partial_r  \left(\frac{\alpha}{a}P + \beta Q \right)
\notag \\
\dot{P} &= \partial_r \left( \frac{\alpha}{a} Q + \beta P \right)   + \frac{\alpha a}{ r^2}w (1 - w^2 )
- g^2 \alpha a w \varphi^2,
\end{align}
and the energy-momentum matter functions reduce to
\begin{align} \label{ADM matter functions}
\rho &=
\frac{\Phi^2 + \Pi^2}{2a^2} 
+ \frac{w^2 \varphi^2}{r^2} + V
+ \frac{(1 - w^2)^2 }{2g^2 r^4}
+ \frac{Q^2+ P^2 }{g^2a^2 r^2}
\notag \\
S\indices{^r_r} &=
\frac{\Phi^2 + \Pi^2}{2a^2} 
- \frac{w^2 \varphi^2}{r^2} - V
- \frac{(1 - w^2)^2 }{2g^2  r^4}
+ \frac{Q^2  + P^2  }{g^2a^2 r^2}
\notag \\
S\indices{^\theta_\theta} 
&= S\indices{^\phi_\phi}   
= \frac{\Pi^2 - \Phi^2}{2a^2}
-V
+ 
 \frac{(1 - w^2)^2 }{2g^2 r^4}
\notag   \\
j_r
&= -\frac{\Phi\Pi}{a}
- \frac{2 Q P }{g^2 a  r^2}.
\end{align}


\subsection{Boundary Conditions}
\label{sec:boundary conditions}

To solve the system of equations I need boundary conditions for many of the variables and, in the case of dynamic solutions, initial data.  Boundary conditions include both conditions at the boundary of space and the boundary of the computational domain.  I list a number of boundary conditions in this subsection, with additional boundary conditions and initial data presented when needed.

The matter part of the monopole is parameterized in terms of the functions $\varphi(t,r)$, representing the scalar field, and $w(t,r)$, representing the gauge field.  If the vacuum value of $\varphi$ is $v$ then their well-known boundary conditions are \cite{Goddard:1977da}
\begin{equation}
\begin{split}
\varphi(t,0) &= 0, \qquad
\varphi(t,\infty) = \pm v,
\\
w(t,0) & = 1, \qquad
w(t,\infty) = 0,
\end{split}
\end{equation} 
with a plus sign for the monopole and a negative sign for the antimonopole.  Their parity properties are $\varphi$ is odd and $w$ is even. 

Inner boundary conditions for metric functions follow from finiteness of the metric equations (\ref{radial-polar metric equations}) and (\ref{radial-maximal metric equations}).  These are $a(t,0) = 1$, which is the flat space value $a$ has when inside a spherically symmetric matter distribution, and $\beta(t,0) = K\indices{^r_r}(t,0) = 0$.  As can be seen by the $\alpha'$ equation in (\ref{radial-polar metric equations}) and the $\alpha''$ equation in (\ref{radial-maximal metric equations}) any solution for $\alpha$ can be scaled by a constant and still be a solution.  Thus one may use $\alpha(t,r)=1/a(t,r)$ for large $r$, which follows from the assumption that the spacetime is asymptotically Reissner-N\"ordstrom.  Parity properties are $a$, $\alpha$, and $K\indices{^r_r}$ are even and $\beta$ is odd.  It follows that $\Phi$ and $P$ are even and $\Pi$, $Q$, and $\alpha'$ are odd.


\section{Static Solutions}
\label{sec:static solutions}

Static monopole solutions in curved space were first studied by van Nieuwenhuizen, Wilkinson, and Perry \cite{VanNieuwenhuizen:1975tc} and later by Lee, Nair, and Weinberg \cite{Lee:1991vy}, Ortiz \cite{Ortiz:1991eu}, and Breitenlohner, Forg\'acs, and Maison \cite{Breitenlohner:1991aa}.  In this section I review only those aspects of static solutions that I need for the next section, where we will find that the static solutions are end states of dynamic solutions and are critical solutions.  A comprehensive analysis of static solutions is given in \cite{Breitenlohner:1991aa, Breitenlohner:1994di}.

A standard approach for finding static solutions is to convert the 't Hooft--Polyakov ansatz,
\begin{equation} \label{monopole ansatz}
\varphi^a = \frac{x^a}{r} \varphi(r), \quad
A_0^a = 0, \quad
A_i^a = -\epsilon_{iak} \frac{x^k}{r} \frac{1 - w(r)}{r},
\end{equation}
written here in Cartesian coordinates, to spherical coordinates, insert it into the flat space Yang-Mills-Higgs Lagrangian, couple the Lagrangian to spherically symmetric gravity, and then derive the equations of motion \cite{VanNieuwenhuizen:1975tc}.  Since I have the complete spherically symmetric time-dependent system of equations, which I listed in the previous section, I shall instead start with them and take the static limit.  As the monopole solution is known to be spherically symmetric and with vanishing electric field, I may use the reduced set of equations in (\ref{reduced evolution equations}).  I note that the gauge field given in (\ref{gauge field}) cannot be directly compared to the ansatz in (\ref{monopole ansatz}), even after (\ref{monopole ansatz}) is converted to spherical coordinates, because they are written in different gauges. Gauge transforming (\ref{monopole ansatz}) with gauge parameter $U = \exp(iT_2 \theta)\exp(iT_3\phi)$ will put (\ref{monopole ansatz}) in the same gauge as (\ref{gauge field}).  The equations of motion can be compared directly without making the gauge transformation.

Static solutions are most easily found in radial-polar gauge.  Upon setting $\Pi = P = 0$ I have four equations:\ the two radial-polar metric equations in (\ref{radial-polar metric equations}) and the $\dot{\Pi}$ and $\dot{P}$ evolution equations in (\ref{reduced evolution equations}) but with their left hand sides set to zero.  It is convenient to parameterize the metric functions in terms of $\sigma(r)$ and the mass function $m(r)$ defined by
\begin{equation}
\begin{split}
\sigma(r) &\equiv \alpha(r) a(r)
\\
N(r) &\equiv 1 - \frac{2Gm(r)}{r} \equiv \frac{1}{a^2(r)},
\end{split}
\end{equation}
where I also introduced $N(r) \equiv 1/g_{rr}$ for convenience.  The resulting system of equations is
\begin{align} \label{static monopole eqs}
\frac{\sigma'}{\sigma} &= 4\pi G \left(r \varphi^{\prime 2} + \frac{2 w^{\prime 2}}{g^2 r} \right) 
\notag \\
m' &= 
4\pi \biggl[
\frac{Nw^{\prime 2}}{g^2} + \frac{(w^2-1)^2}{2g^2 r^2} + \frac{r^2}{2} N \varphi^{\prime 2}
\notag \\ 
&\qquad\qquad
+ w^2 \varphi^2 + r^2 V \biggr] 
\notag \\
\partial_r \left(r^2 N \sigma \varphi' \right)
&= \sigma \left(2w^2\varphi + r^2 \frac{\partial V}{\partial \varphi} \right) 
\notag \\
\partial_r \left(N \sigma w' \right) &= \sigma \left[ \frac{w(w^2 - 1)}{r^2} + g^2 \varphi^2 w \right].
\end{align}

In the literature there exist two common mass scales used for constructing dimensionless quantities:\ $m_P$ and $v$, where $m_P = 1/\sqrt{G}$ is the Planck mass and $v$ is the vacuum value of the scalar field.  
I shall use the mass scale $m_P$ and thus introduce
\begin{gather}
\bar{r} \equiv (g m_G) \hspace{0.1 em} r, \quad
\bar{\varphi} \equiv \varphi/m_G, \quad
\bar{m} \equiv (g m_G/m_P^2) \hspace{0.1 em} m, 
\notag \\
\bar{v} \equiv v/m_G, \quad
\bar{\lambda} \equiv \lambda/g^2, 
 \label{scaling}
\end{gather}
where $m_G \equiv m_P / \sqrt{4\pi}$ with the $\sqrt{4\pi}$ included for convenience.
I note that $w$ is already dimensionless and $\bar{v} = m_V/gm_G$ and $\bar{\lambda} = (m_S /\sqrt{2} m_V)^2$, where $m_V$ and $m_S$ are the vector and scalar masses in (\ref{vector scalar mass}).  

The $\sigma'$ equation in (\ref{static monopole eqs}) may be used to eliminate $\sigma$ from the other equations after which it decouples.  Since I do not need the result for $\sigma$ in radial-polar gauge I drop the $\sigma'$ equation.  Moving to dimensionless quantities the system of equations becomes
\begin{equation} \label{static equations}
\begin{split}
\bar{m}' &= 
N w^{\prime 2} + \frac{(w^2-1)^2}{2 \bar{r}^2} + \frac{1}{2} \bar{r}^2 N \bar{\varphi}^{\prime 2} 
\\
&\qquad
+ w^2 \bar{\varphi}^2
 + \bar{r}^2 \frac{\bar{\lambda}}{4} \left(\bar{\varphi}^2 - \bar{v}^2 \right)^2
\\
\bar{\varphi}'' & = -\bar{\varphi}'
\left(  \frac{2}{\bar{r}} + \frac{N'}{N}  +  \bar{r} \bar{\varphi}^{\prime 2} + \frac{2 w^{\prime 2}}{\bar{r}} \right)
\\
&\qquad + \frac{1}{N} \left[ \frac{2 w^2 \bar{\varphi}}{\bar{r}^2} + \bar{\lambda} \bar{\varphi}\left(\bar{\varphi}^2 - \bar{v}^2 \right) \right]
\\
w'' & =-  w' \left( \frac{N'}{N} +  \bar{r} \bar{\varphi}^{\prime 2} + \frac{2 w^{\prime 2}}{ \bar{r}} \right)
\\
&\qquad + \frac{1}{N} \left[ \frac{w(w^2-1)}{\bar{r}^2} + \bar{\varphi}^2 w\right],
\end{split}
\end{equation}
where a prime indicates a derivative with respect to $\bar{r}$ and where
\begin{equation}
N = 1 - \frac{2\bar{m}}{\bar{r}}, \qquad
N' = \frac{2\bar{m}}{\bar{r}^2} - \frac{2}{\bar{r}}\bar{m}'.
\end{equation}
This system of equations has regular and singular (black hole) solutions.  Once appropriate boundary conditions are identified the equations can be solved numerically using standard integration techniques.  

To reduce the vast number of solutions that can be presented in this and the next section I consider only fundamental monopoles and ignore the excited solutions (i.e.\ solutions with a nonzero number of nodes or zero-crossings of the gauge field).  I also restrict attention to $\lambda = 0$ and $\bar{v} = 0.2$, 0.3, and 0.4.  At the end I'll comment on results for nonzero $\lambda$ and other values of $\bar{v}$.

\begin{figure}
\centering
\includegraphics[width=3.2in]{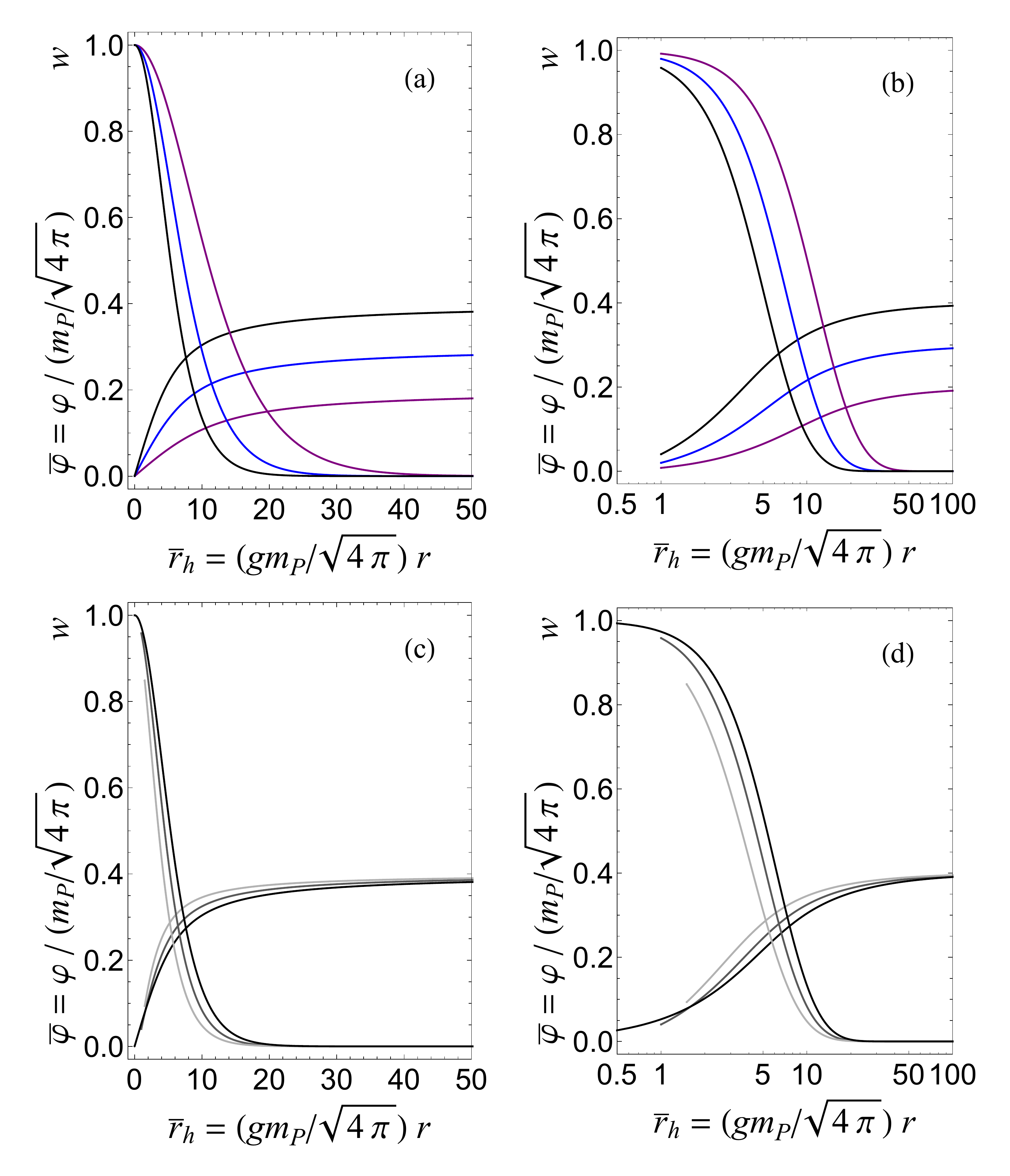}
\caption{(a) Regular monopole solutions for $\lambda = 0$ and (from right to left for $w$ and bottom to top for $\varphi$) $\bar{v} = 0.2$ (purple), 0.3 (blue), and 0.4 (black).  (b) Black hole monopole solutions for $\lambda = 0$, $\bar{r}_h = 1$, and the same values of $\bar{v}$ as in (a).  (c) and (d) are the same plot with (d) on a log scale to see more easily the region just outside the black hole.  Both show regular and black hole monopole solutions for $\lambda = 0$, $\bar{v} = 0.4$, and $\bar{r}_h = 0$ (black), 1 (medium gray), and 1.5 (light gray).}
\label{figure1}
\end{figure}

For regular monopoles the boundary conditions were given in Sec.\ \ref{sec:boundary conditions}.  In particular $\varphi$ is odd and $\varphi(0) = 0$, $w$ is even and $w(0) = 1$, and $m$ is odd and $m(0) = 0$.  After expanding these quantities in a power series consistent with these properties and plugging them into the system of equations in (\ref{static equations}) it can be shown that near the origin \cite{Lee:1991vy, Ortiz:1991eu, Breitenlohner:1991aa}
\begin{equation}
\bar{m} = O(r^3), \quad 
\bar{\varphi} = c\bar{r} + O(r^3), \quad 
w = 1 - b\bar{r}^2 + O(r^4). 
\end{equation}
Given values for $b$ and $c$ the above equations give inner values at some small $\bar{r} = \bar{r}_\text{min}$, from which solutions for $\bar{r} > \bar{r}_\text{min}$ can be found by integrating outward.  The constants $b$ and $c$ are determined using the shooting method, with outer boundary conditions $\bar{\varphi} = \pm \bar{v}$ and $w = 0$ at large $r$.  In this and the next section I consider only $\bar{\varphi} = +\bar{v}$ solutions.  Once a solution is found the asymptotic value of $\bar{m}$ is the ADM mass $\overline{M} = (g/\sqrt{4\pi} m_P) M$.  Solutions are shown in Fig.\ \ref{figure1}(a).

The outer boundary conditions for black hole monopoles are the same as for regular monopoles since in both cases the spacetime is assumed asymptotically flat.  The inner boundary conditions are $N(\bar{r}_h) = 0$ and $N(\bar{r}) > 0$ for $\bar{r} > \bar{r}_h$, where $\bar{r}_h$ is the horizon radius.  As with the regular solutions we expand the quantities, but this time around $\bar{r}_h$, and plug them into (\ref{static equations}).  Since $\varphi$ and $w$ should be regular across the horizon the coefficients of terms that go like $1/(\bar{r}-\bar{r}_h)$ must vanish.  The result is \cite{Breitenlohner:1991aa}
\begin{align}
\bar{\varphi} &= \bar{\varphi}_h
+
\frac{1}{N_1} \left[\frac{2 w_h^2 \bar{\varphi}_h}{\bar{r}_h^2} + \bar{\lambda} \bar{\varphi}_h(\bar{\varphi}_h^2-\bar{v}^2) \right] x
+O(x^2) 
\notag \\
w &= w_h +  \frac{w_h}{N_1} \left( \frac{w_h^2 - 1}{\bar{r}_h^2} + \bar{\varphi}_h^2 \right) x
+ O(x^2) 
\notag \\
N &= N_1 x + O(x^2)
\end{align}
where $x \equiv \bar{r} - \bar{r}_h$ and
\begin{equation}
N_1 = \frac{1}{\bar{r}_h} - \frac{2w_h^2 \bar{\varphi}_h^2}{\bar{r}_h} - \frac{(w_h^2 - 1)^2}{\bar{r}_h^3}
- \bar{r}_h\frac{\bar{\lambda}}{2} (\bar{\varphi}_h^2 - \bar{v}^2)^2 .
\end{equation}
These equations give inner values at $\bar{r}$ very near $\bar{r}_h$, from which the solutions for $\bar{r}>\bar{r}_h$ can be solved for by integrating outward.  As explained in \cite{Breitenlohner:1991aa} black hole monopole solutions for a given $\bar{v}$ are uniquely identified by their value of $w_h$ (or $\bar{\varphi}_h$), but not by $\bar{r}_h$, as there can exist multiple solutions with the same horizon radius.  In practice I fix $w_h$ to any value in $0 < w_h < 1$ and use the shooting method to determine $\bar{\varphi}_h$ and $\bar{r}_h$ for outer boundary conditions $\bar{\varphi} = \bar{v}$ and $w = 0$ at large $r$.  Black hole solutions are shown in Fig.\ \ref{figure1}(b) for $\bar{r}_h = 1$.  Figures \ref{figure1}(c) and (d) are the same plot, with (d) on a log scale, of both regular and black hole solutions with $\bar{v} = 0.4$.  The log scale allows the region just outside the black hole to be seen more easily.  For a comprehensive display of regular and black hole solutions see \cite{Breitenlohner:1991aa}.

The equations in (\ref{static equations}) also contain the Reissner-N\"ordstrom (RN) black hole, which occurs for $\bar{\varphi}$ and $w$ having constant values
\begin{equation} \label{RN BH sol}
\bar{\varphi} = \pm \bar{v}, \qquad
w = 0.
\end{equation}
Moving back to unbarred quantities, the solution is $m = M - (2\pi/g^2)/r$ and thus
\begin{equation}
g_{rr}^{-1} = N = 1 - \frac{2M G}{r} + \frac{4\pi G/g^2}{r^2}.
\end{equation}
As I have set the electric field to zero through the magnetic ansatz this is the RN spacetime with unit magnetic charge $1/g$.  In fact, it can be shown that in the large $r$ limit the general solution to (\ref{static equations}) is the RN solution and thus, in this sense, all monopole solutions also have unit charge \cite{Aichelburg:1992st}.  An RN black hole with unit charge can only exist for $\overline{M} = (g/\sqrt{4\pi} m_P) M \geq 1$ and $\bar{r}_h \geq 1$, where $\bar{r}_h$ is the outer horizon radius.  If $\overline{M} = 1$ it is an extremal RN black hole with a single horizon with radius $\bar{r}_h = 1$.

\begin{figure*}
\centering
\includegraphics[width = 6.4in]{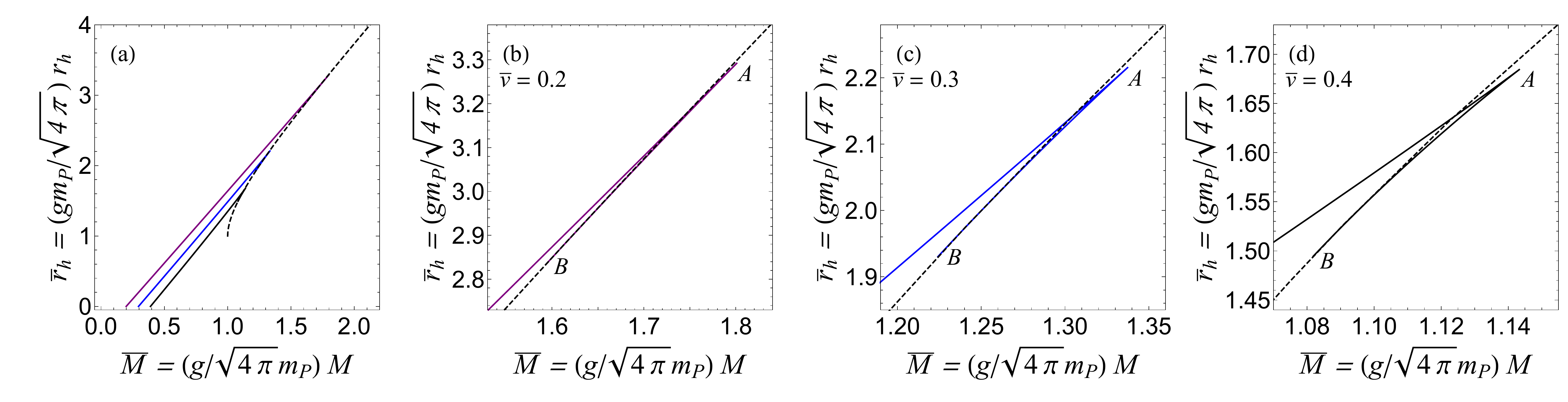}
\caption{Horizon radius $\bar{r}_h$ as a function of mass $\overline{M}$ for static solutions.  In (a) the solid lines from top to bottom are for $\bar{v} = 0.2$ (purple), 0.3 (blue), and 0.4 (black).  The dashed line is the outer horizon of the Reissner-Nordstr\"om (RN) black hole which only exists for $\overline{M} \geq 1$ and $\bar{r}_h \geq 1$.  (b--d) are magnifications of the regions where three static solutions exist with the same mass.  In each $A$ marks the cusp where the two monopole solution branches meet and $B$ indicates the bifurcation where the two branches first appear and specifically marks the edge of the bottom branch.  The values of $(\overline{M}_A, \overline{M}_B)$ for $v = 0.2$ are (1.802, 1.593), for $v = 0.3$ are (1.338, 1.224), and for $v = 0.4$ are (1.144, 1.082).}   
\label{figure2}
\end{figure*}

To recap, the solutions we've found to the system of equations in (\ref{static equations}) include a regular monopole, black hole monopoles, of which there can be multiple solutions with the same horizon radius, and the RN black hole.  In this and the following section I take as the black hole mass the ADM mass, $\overline{M} = \bar{m}(\infty)$.  In Fig.\ \ref{figure2} I've plotted the horizon radius $\bar{r}_h$ of static solutions as a function of their mass $\overline{M}$.  In Fig.\ \ref{figure2}(a) we see that for  $\overline{M} < 1$ there exists a unique static solution for a given $\bar{v}$.  For $\overline{M} \geq 1$ multiple solutions appear.  In the magnifications in Figs.\ \ref{figure2}(b--d) we can see regions where up to three different static black hole solutions have the same mass.  The points labeled $B$ are known as bifurcations, where two monopole solution branches appear, and the points labeled $A$ are known as cusps, where the two branches meet.

An important question is whether these static solutions are stable under (spherically symmetric) radial perturbations?  Lee, Nair, and Weinberg studied the stability of the RN black hole in this system \cite{Lee:1991qs} and found it could be unstable but did not give precise details of its instability with respect to the other solutions.  Aichelburg and Bizo\'n studied the stability of the black hole solutions \cite{Aichelburg:1992st}, but only rigorously for $\lambda \rightarrow \infty$, which effectively fixes $\bar{\varphi} = \bar{v}$ and simplifies the analysis, where they found that (fundamental) monopole black holes are always stable.  Nevertheless, from their results they inferred for finite $\lambda$ that the black hole monopole solutions in the upper branches in Figs.\ \ref{figure2}(b--d) are stable, the solutions in the bottom branches are unstable, and the RN solution is unstable for $1 < \overline{M} < \overline{M}_B$ and stable for $\overline{M} > \overline{M}_B$, where $B$ is the bifurcation point.  Their inference was corroborated by Maeda, Tachizawa, et.\ al.\ \cite{Maeda:1993ap, Tachizawa:1994wn} using catastrophe theory.  Finally, Hollmann studied the stability of regular monopoles \cite{Hollmann:1994fm} and found that for the values of $\bar{v}$ used here they are always stable.  (For larger values of $\bar{v}$ there can exist two regular solutions with the same mass and only the smaller mass solution is stable).  

In Sec.\ \ref{sec:stability} I study stability by dynamically solving the system, allowing for collapse, and determining the end state.  My results corroborate those above.  As far as I am aware, the number of unstable modes in a lower branch solution has not been determined.  If there is only a single unstable (radial) mode then the solution is a prime candidate for being a critical solution.  In Sec.\ \ref{sec:critical} I find strong evidence that the unstable lower branch solutions are in fact critical solutions.


\section{Dynamic Solutions}
\label{sec:dynamic solutions}

In this section I present the principle results of this paper, the dynamic evolution of spherically symmetric spacetimes containing gravitational monopoles.  I wish to determine the final state of the evolution and thus need code that retains stability well after black hole formation.  For this reason I use black hole excision methods.  The black hole excision methods, boundary conditions, and initial data I make use of are presented in the next subsection.  In subsequent subsections I present results for stability and critical behavior and comment on the no-hair conjecture for this model.


\subsection{Black Hole Excision, Boundary Conditions, and Initial Data}
\label{sec:BH excision}

The matter fields to solve for are $\varphi(t,r)$ for the scalar field and $w(t,r)$ for the gauge field.  They obey the (time-dependent) evolution equations in (\ref{reduced evolution equations}).  The metric functions to solve for in radial-maximal gauge are $a(t,r)$, $\alpha(t,r)$, and $K\indices{^r_r}(t,r)$, where I use $K\indices{^r_r}$ instead of $\beta$ through (\ref{beta Krr}).  They obey the (time-dependent) constraint equations in (\ref{radial-maximal metric equations}).  To find numerical solutions I put the equations in first order form and scale the quantities.  First order form requires only the introduction of the equation $\alpha' = \delta$ and all other occurrences of $\alpha'$ to be replaced with $\delta$.  For scaling I again use (\ref{scaling}) along with $\bar{t} \equiv (gm_G) t$ and $\overline{K}\indices{^r_r} \equiv K\indices{^r_r}/gm_G$.

The constraint equations in (\ref{radial-maximal metric equations}) determine the metric functions on a single time-slice as long as boundary conditions are available.  In the absence of a black hole the origin has not been excised and the boundary conditions are as given in Sec.\ \ref{sec:boundary conditions}.  I determine if a black hole has formed by searching for an apparent horizon.  In spherical symmetry this is straightforward and I simply check on each time-slice if (\ref{apparent horizon condition}) is satisfied.  If an apparent horizon is found, from that time forward I excise all grid points that satisfy (\ref{apparent horizon condition}).  I continue to use the equations in (\ref{radial-maximal metric equations}) to determine the metric functions, but now since the origin has been excised I can no longer use the inner boundary conditions in Sec.\ \ref{sec:boundary conditions}.  Following \cite{Choptuik:1999gh} I find the inner boundary values by ``freezing" $\alpha$ and $\alpha'$ at the time-step directly before excision at what will become the new inner boundary and using the evolution equations for $a$ and $K\indices{^r_r}$ in (\ref{a Krr evolution equations}).

Finally I need outer boundary conditions for the matter functions.  Since the computational domain does not extend to $r = \infty$ I must allow the matter fields to be able to exit the computational domain.  I use standard outgoing wave and radiation conditions.  At the outer boundary I approximate the spacetime as flat and in the large $r$ limit so that $\alpha = a = 1$ and $\beta = K\indices{^r_r} = \partial_\varphi V = 0$, reducing the matter evolution equations in (\ref{reduced evolution equations}) to
\begin{equation}
(-\partial_t^2 + \partial_r^2)(r\varphi) = 0, \qquad
(-\partial_t^2 + \partial_r^2)w = 0,
\end{equation}
which is to say $\varphi$ is a spherical wave and $w$ is a one-dimensional wave.  The standard technique is to assume that both $\varphi$ and $w$ can only be outgoing waves at the outer boundary and thus I am ignoring backscattering caused by the curvature of spacetime there.  If I make the computational grid large enough this should be a reasonable approximation.  I thus assume $\varphi = \pm v + f_\varphi(r-t)/r$ for some function $f_\varphi$, i.e.\ that it has the form of an outgoing spherical wave, and $w = f_w(r-t)$ for some function $f_w$.  Note that $\Pi$, $Q$, and $P$ must also have outgoing wave forms but $\Phi$ cannot.  It follows that the outer boundary conditions are
\begin{equation}
\begin{split}
\dot{\varphi} &= - (\varphi \mp v)/r - \Phi \\
\dot{\Pi} &= - \Pi/r - \Pi' \\
\Phi &= -(\varphi \mp v)/r - \Pi \\
\dot{w} &= - Q \\
\dot{P} &= Q' \\
Q &= -P.
\end{split}
\end{equation}

For initial data I adapt that used in \cite{Choptuik:1999gh, Millward:2002pk} to the monopole system:
\begin{align} \label{initial data}
\varphi(0,r) &= v \tanh \left( \frac{r}{s_\varphi} \right)
\notag \\
w(0,r) &=\frac{1}{2}\Biggl\{ 1 + \ \left[1 + a_w \left(1 + \frac{b_w r}{s_w}\right) e^{-2(r/s_w)^2} \right]
\notag \\
&\qquad \times \tanh \left( \frac{x_w-r}{s_w}\right)
\Biggr\}
\end{align}
and $\dot{\varphi}(0,r) = \dot{w}(0,r) = 0$, making it time-symmetric.  The parameters $x_w$ and $s_w$ give the center and spread of the $w$-pulse and the parameters $a_w$ and $b_w$ are chosen such that the gauge field boundary conditions are satisfied at the origin and are given by
\begin{equation}
a_w = \coth(x_w/s_w) - 1, \quad b_w = \coth(x_w/s_w)+1.
\end{equation}
 
I composed second order accurate code to dynamically evolve the spacetime, which was inspired by the description Choptuik, Hirschmann, and Marsa gave of their code in \cite{Choptuik:1999gh} (see also \cite{Millward:2002pk}).  I use a staggered grid that does not include the origin with virtual grid points at each boundary so that in the absence of a black hole all spatial derivatives in the computational domain can be finite-differenced with centered stencils.  Inner virtual grid points also serve to impose the parity properties of the fields.  In the presence of a black hole one-sided stencils are used near the apparent horizon/inner boundary.  I solve all constraint equations using second order Runge-Kutta and all evolution equations using the method of lines and third order Runge-Kutta.  All results are made with $\bar{r}_\text{max} = 100.005$, $\Delta\bar{r} = 0.01$ (unless stated otherwise), $\Delta \bar{t}/\Delta \bar{r} = 0.5$, and for determining the apparent horizon $\mu_H = 1.1$.  

\begin{figure}
\centering
\includegraphics[width=3in]{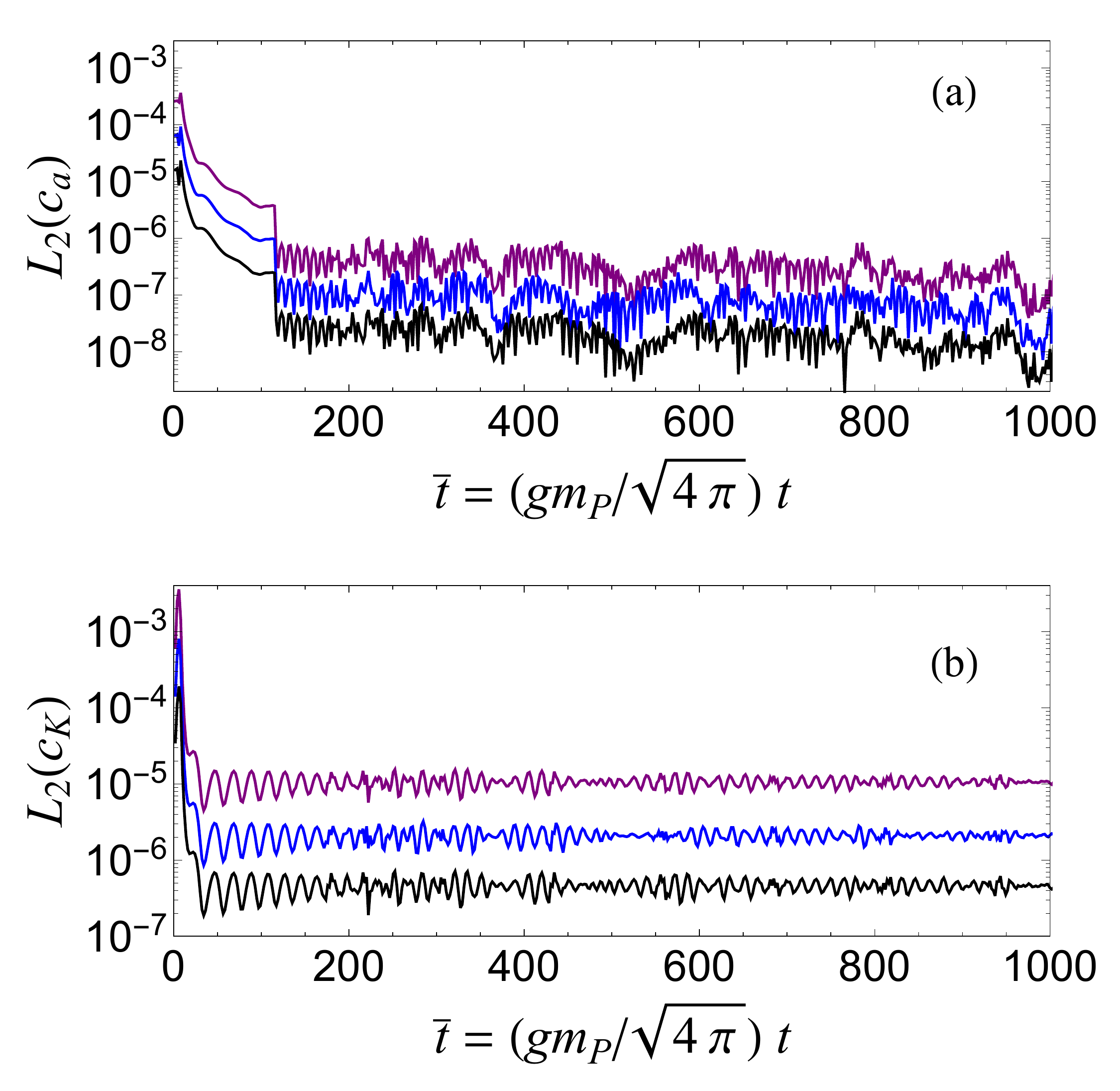}
\caption{The $L_2$ norm across the computational grid for $c_a = \dot{a} - (\cdots)$ and $c_K = \dot{K}\indices{^r_r} - (\dots)$, where the dots represent the respective right hand sides of (\ref{a Krr evolution equations}).  The results are shown for three spatial resolutions:\ (from top to bottom) $\Delta r = 0.02$ (purple), 0.01 (blue), and $0.005$ (black).  That the results are small indicates that the constraints $c_a = 0$ and $c_K = 0$ are obeyed and that the results drop by a factor of 4 (for spatial resolutions that drop by a factor of 2) indicates second order convergence.}
\label{fig:convergence}
\end{figure}

The evolution equations for $a$ and $K\indices{^r_r}$ in (\ref{a Krr evolution equations}) are only used at a single grid point, and only if a black hole forms, and thus are available for consistency checks on the code.  Defining $c_a \equiv \dot{a} - (\cdots)$ and $c_K \equiv \dot{K}\indices{^r_r} - (\cdots)$, where the dots represent the respective right hand sides of (\ref{a Krr evolution equations}), I've plotted the $L_2$ norm of $c_a$ and $c_K$ across the computational domain in Fig.\ \ref{fig:convergence} for three spatial resolutions:\ $\Delta \bar{r} = 0.02$, 0.01, and $0.005$.  The results shown are for the same initial data used below in Fig.\ \ref{fig:regular monopole}, but I've found them to be typical, including for when a black hole forms.  That the results are small indicates that the constraints $c_a = 0$ and $c_K = 0$ are obeyed and that the results drop by a factor of 4 (for spatial resolutions that drop by a factor of 2) indicates second order convergence.


\subsection{End States and Stability}
\label{sec:stability}

I study stability of the static solutions reviewed in Sec.\ \ref{sec:static solutions} by evolving initial data until a final stable configuration is reached.  The initial data require specification of $\bar{\lambda}, \bar{v}, \bar{x}_w, \bar{s}_w$, and $\bar{s}_\varphi$.  I mentioned in the previous section that I'll set $\bar{\lambda} = 0$ and focus on $\bar{v} = 0.2$, 0.3, and 0.4.  I've tried various values of $\bar{s}_\varphi$ and will fix $\bar{s}_\varphi = 10$ \cite{Millward:2002pk} which gives typical results.  This leaves $\bar{x}_w$ and $\bar{s}_w$ which will be free parameters I search through.

\begin{figure*}[t]
\centering
\includegraphics[width=6in]{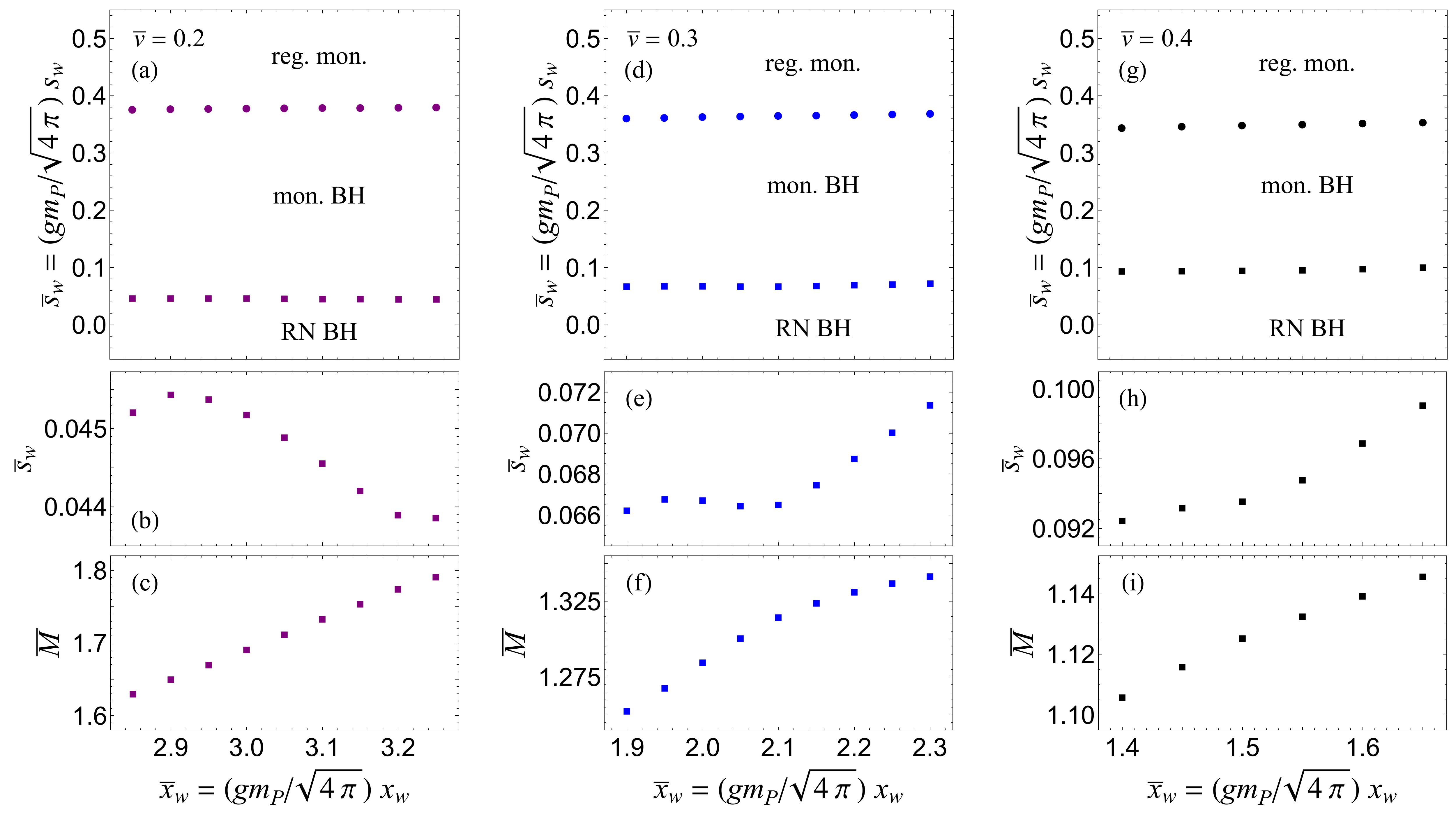}
\caption{``Phase" or ``end state" diagrams indicating the stable end state (regular monopole, monopole black hole, or RN black hole) of an evolution  starting from initial data (\ref{initial data}) with $\bar{\lambda} = 0$ and $\bar{s}_\varphi = 10$.  Each column is for the value of $\bar{v}$ indicated.  Circles mark the threshold of collapse and squares mark the transition between monopole and RN black holes.  The middle row zooms in on the squares and the bottom row plots the squares as a function of mass $\overline{M}$ (with $\bar{s}_w$ suppressed).}
\label{fig:end state}
\end{figure*}

For an initial sense of the system let's begin with the ``phase" or ``end state" diagrams in Fig.\ \ref{fig:end state}.
The diagrams indicate the stable end state of an evolution beginning with initial data (\ref{initial data}).  Each column is for a different value of $\bar{v}$.  Consider first Fig.\ \ref{fig:end state}(a) for $\bar{v} = 0.2$.  Every evolution I tried always ended in one of three end states:\ the regular monopole, a monopole black hole, or the RN black hole.  This is not surprising as these are the only three static solutions found in Sec.\ \ref{sec:static solutions}.  The circles mark the threshold of collapse and the squares mark the transition between monopole and RN black holes and both were found by fixing $\bar{x}_w$ and searching through $\bar{s}_w$.  In the next subsection I focus on the monopole and RN black hole transition and for this reason I zoom in on the squares in Fig.\ \ref{fig:end state}(b) to show their variation.  Further, the mass $\overline{M}$ of the stable end state will be important for that analysis and in Fig.\ \ref{fig:end state}(c) I again plot the squares but in terms of $\overline{M}$ (with $\bar{s}_w$ suppressed).  The other two columns are the same but for $\bar{v} = 0.3$ and $0.4$.  The choice of which $\bar{x}_w$ values to show is explained by the bottom row, in that these values of $\bar{x}_w$ lead to black holes (at the threshold between monopole and RN black holes) with masses that nicely span the masses at which all three black holes in Figs.\ \ref{figure2}(b--c) coexist.

The top row of Fig.\ \ref{fig:end state} indicates that for large $\bar{s}_w$ there is no collapse and the final state of the system is the regular monopole.  This is reasonable since as $\bar{s}_w$ increases energy in the initial $w$-pulse spreads out and collapse becomes less likely.  For the initial data used I find that $\bar{v}$ and $\bar{x}_w$ do not have much effect on the onset of collapse (for different initial data, of course, this is not necessarily the case).  For middle values of $\bar{s}_w$ collapse occurs and the final states are black hole monopoles while for sufficiently small $\bar{s}_w$ the final states are RN black holes.  The RN black hole requires small $\bar{s}_w$ because the black hole that forms must be large enough to, in a sense, ``swallow the monopole" \cite{Volkov:1998cc}.

Now let's take a closer look at the three possible end states. 
\begin{figure*}
\centering
\includegraphics[width=6.5in]{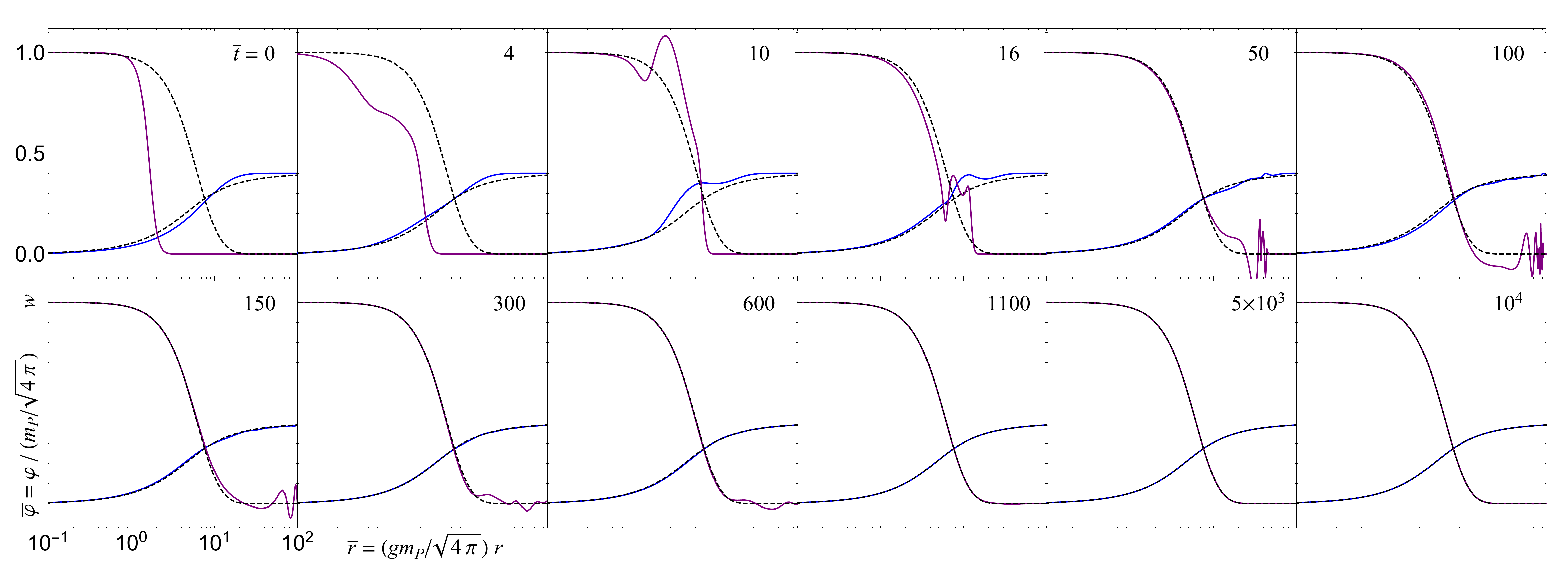}
\caption{Typical time evolution of the scalar field $\bar{\varphi}$ (blue) and gauge field $w$ (purple) for initial data (\ref{initial data}) in which the end state is the static regular solution (dashed lines).  The initial data use $(\bar{v}, \bar{\lambda}, x_w, s_w, s_\varphi) = (0.4, 0, 1.6, 0.4, 10)$.  All frames have the same axes.  The coordinate time $\bar{t}$ for each frame is given in the corner.}
\label{fig:regular monopole}
\end{figure*}
\begin{figure*}
\centering
\includegraphics[width=6.5in]{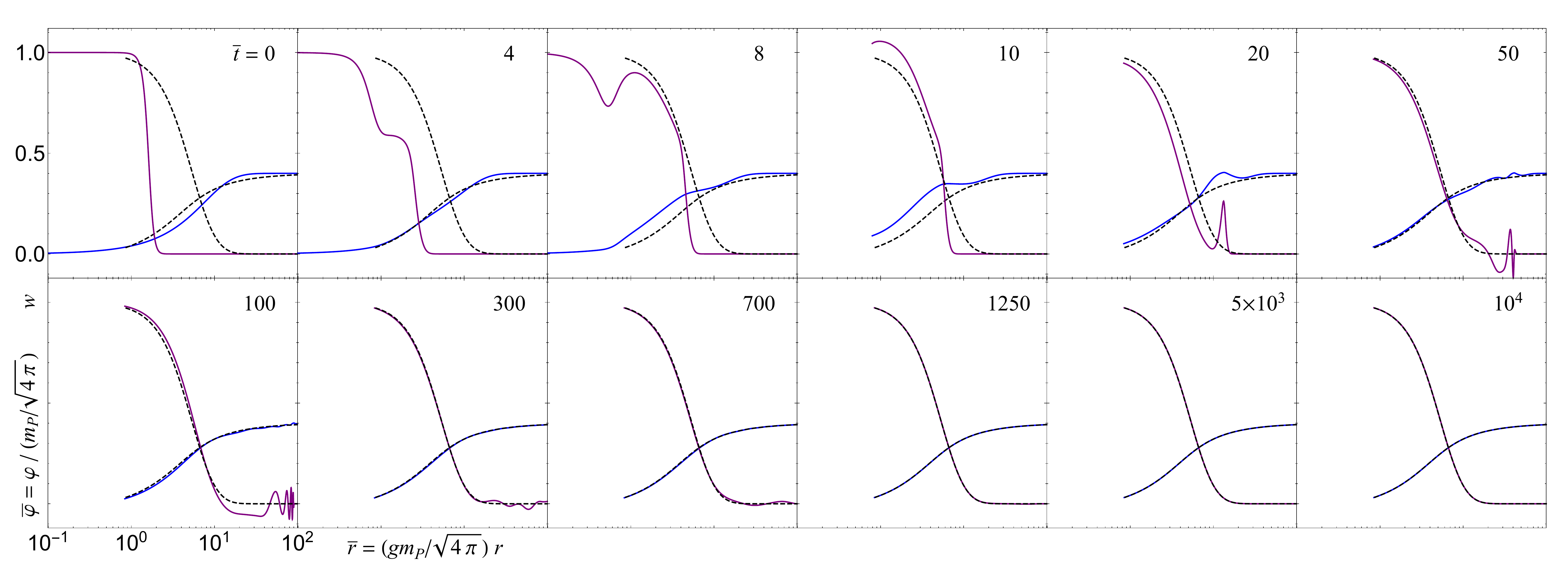}
\caption{The same as Fig.\ \ref{fig:regular monopole} except with the end state being a black hole monopole.  The initial data use $(\bar{v}, \bar{\lambda}, x_w, s_w, s_\varphi) = (0.4, 0, 1.6, 0.25, 10)$ and the end state has mass $\overline{M} = 0.7856$.}
\label{fig:monopole BH}
\end{figure*}
\begin{figure*}
\centering
\includegraphics[width=6.5in]{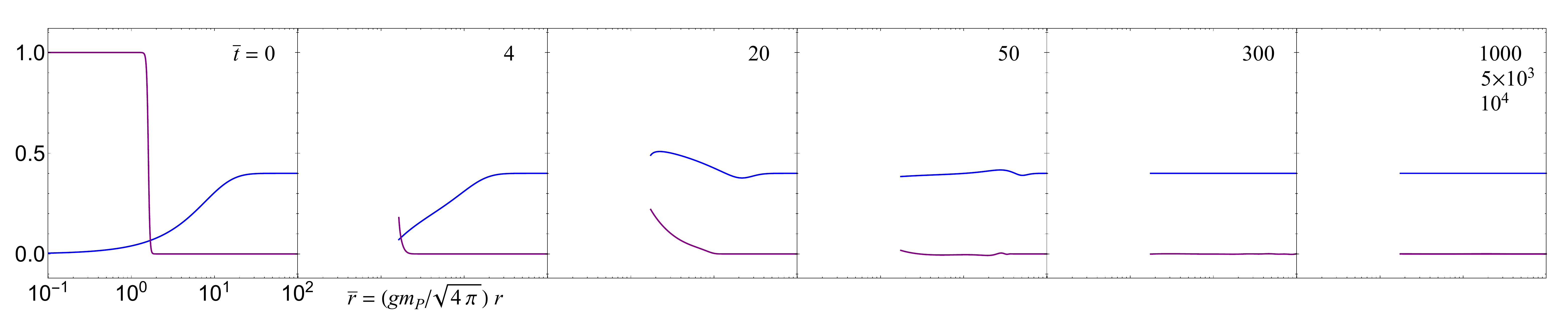}
\caption{The same as Figs.\ \ref{fig:regular monopole} and \ref{fig:monopole BH} except with the end state being the RN black hole (see (\ref{RN BH sol})).  The initial data use $(\bar{v}, \bar{\lambda}, x_w, s_w, s_\varphi) = (0.4, 0, 1.6, 0.07, 10)$ and the end state has mass $\overline{M} = 1.181$.  Solutions at three different times are plotted in the final frame.}
\label{fig:RN BH}
\end{figure*}
Figures\ \ref{fig:regular monopole}, \ref{fig:monopole BH}, and \ref{fig:RN BH} show typical evolutions in which the end states are, respectively, a regular monopole, a monopole black hole, and an RN black hole.  In Fig.\ \ref{fig:regular monopole} the dashed lines show the static regular monopole solution from Sec.\ \ref{sec:static solutions}, which the dynamic solution is seen to evolve to.  Similarly in Fig.\ \ref{fig:monopole BH} the dashed lines are a static black hole monopole solution.  In Fig.\ \ref{fig:RN BH} the dynamic solution is seen to evolve to the RN solution (\ref{RN BH sol}). 

For $\bar{v} = 0.2$, 0.3, or 0.4 there is a unique regular monopole solution and comparing it with the end state of the dynamic evolution is trivial.  To make the analogous comparison when the end state is a monopole black hole I match masses at the outer boundary of the computational domain, that is I take the value of $\bar{m}(\bar{t},\bar{r}_\text{max})= \bar{r}_\text{max}(1-1/a^2(\bar{t},\bar{r}_\text{max}))/2$ of the end state, use results similar to those used to make Fig.\ \ref{figure2} to find the value of $w_h$ for a static solution with the same $\bar{m}(\bar{r}_\text{max})$, construct the static solution, and compare.  (In Sec.\ \ref{sec:static solutions} I noted that at large $r$ the general solution is the RN solution and thus from $\bar{m}(\bar{t},\bar{r}_\text{max})$ I can infer $\overline{M}$, which differs by only a small amount and is what I plotted in the bottom row of Fig.\ \ref{fig:end state}).

\begin{figure}
\centering
\includegraphics[width=3.2in]{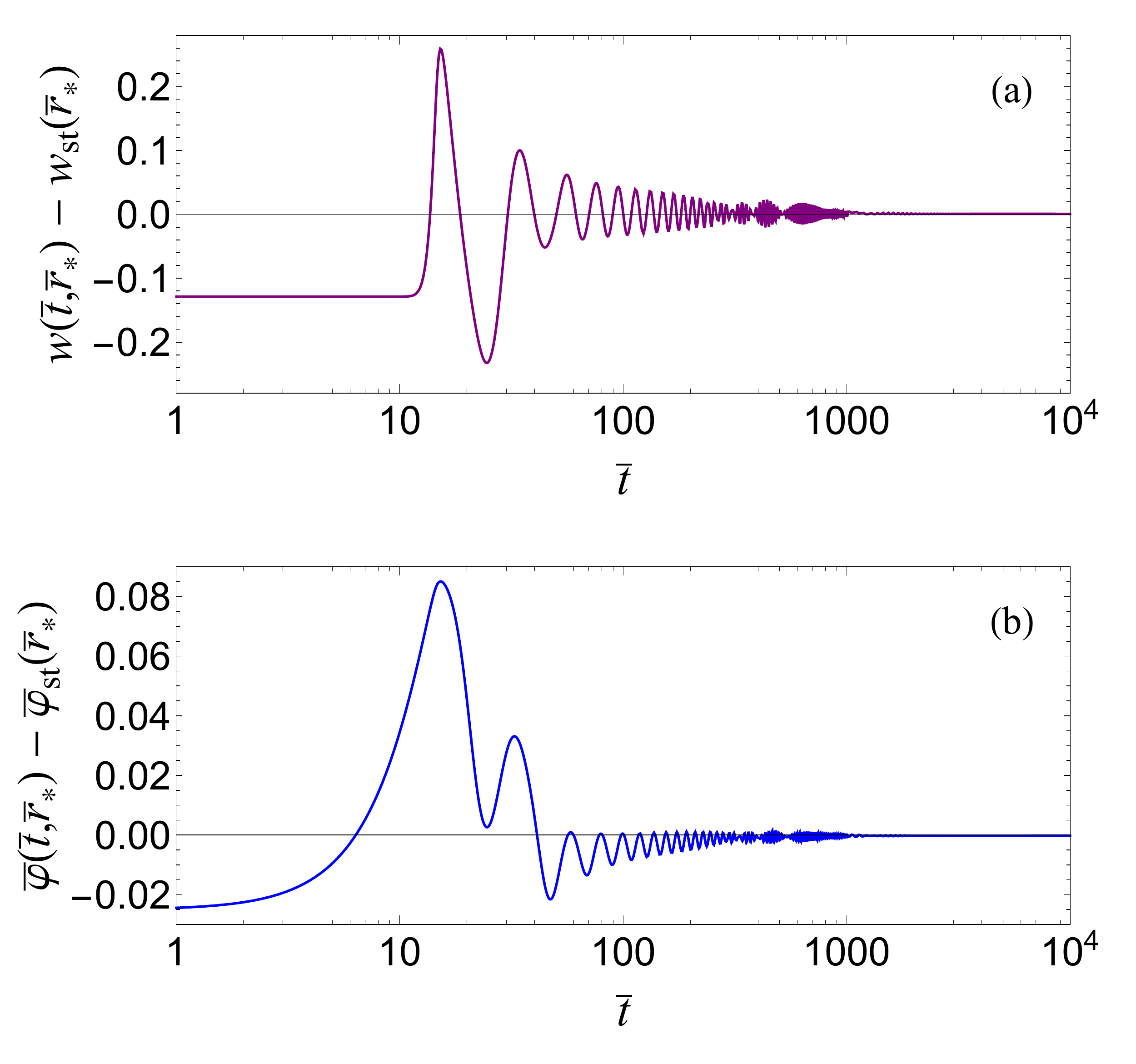}
\caption{$w(\bar{t},\bar{r}_*)$ and $\bar{\varphi}(\bar{t},\bar{r}_*)$ are the same dynamic solutions shown in Fig.\ \ref{fig:monopole BH}, with $w_\text{st}$ and $\bar{\varphi}_\text{st}$ the corresponding static solutions.  These plots are for $\bar{r}_* = 9.005$ where $w_\text{st}(\bar{r}_*) = 0.1289$ and $\bar{\varphi}_\text{st}(\bar{r}_*) = 0.3117$.  Analogous plots for different values of $\bar{r}_*$ and different evolutions look very similar.}
\label{fig:ringdown}
\end{figure}

Another way to view the evolution of the system toward the static solution is shown in Fig.\ \ref{fig:ringdown}, which displays $w(\bar{t},\bar{r}_*) - w_\text{st}(\bar{r}_*)$ and $\bar{\varphi}(\bar{t},\bar{r}_*) - \bar{\varphi}_\text{st}(\bar{r}_*)$, where $w$ and $\bar{\varphi}$ are the same dynamic solutions shown in Fig.\ \ref{fig:monopole BH} and $w_\text{st}$ and $\bar{\varphi}_\text{st}$ are the corresponding static solutions (dashed curves in Fig.\ \ref{fig:monopole BH}).   Figure \ref{fig:ringdown} is for $\bar{r}_* = 9.005$, but other values of $\bar{r}_*$ and other evolutions (for example those in Figs.\ \ref{fig:monopole BH} and \ref{fig:RN BH}) give very similar looking results.

For all initial data I tried if a black hole does not form (specifically if an apparent horizon is never found) I always found the end state to be the regular monopole.  This is strong support that these regular monopole solutions are stable.  If a black hole forms and the mass of the end state satisfies $\overline{M} < \overline{M}_B$, where $B$ is the bifurcation point in Fig.\ \ref{figure2}(b--d), I have only been able to find monopole black holes and have never found an RN black hole, suggesting in this region the monopole black hole is stable and the RN black hole is unstable.  For end states with $\overline{M} > \overline{M}_A$, where $A$ marks the cusp in Figs.\ \ref{figure2}(b-d), I have only found RN black holes, suggesting that in this region RN black holes are stable and monopole black holes cannot form.  All of this corroborates the stability discussion in Sec.\ \ref{sec:static solutions}.  Of course the most interesting region is $\overline{M}_B < \overline{M} < \overline{M}_A$ where three static black hole solutions coexist.  I study this region in the next subsection.


\subsection{Critical Behavior and Stability}
\label{sec:critical}

Critical behavior in gravitational collapse was first discovered by Choptuik \cite{Choptuik:1992jv}.  In gravitational systems with one-parameter families of initial data from which collapse can occur there exists a critical value of the parameter such that, say, above the critical value, $p > p^*$, collapse does not occur and below the critical value, $p<p^*$, collapse occurs.  The resulting spacetime for $p = p^*$ is the critical solution. 

Critical solutions are attractors, but contain a single decay mode and are unstable.  This means for $p$ sufficiently close to $p^*$ the spacetime will evolve to be very close to the critical solution before moving away to either a spacetime with a black hole or one without.  The closer $p$ is to $p^*$ the longer the spacetime stays near the critical solution before decaying.  Choptuik discovered type II critical behavior in \cite{Choptuik:1992jv} and subsequently Choptuik, Chmaj, and Bizo\'n discovered type I critical behavior in \cite{Choptuik:1996yg}.  In type I the critical solution is a stationary (or periodic) spacetime and black holes form with finite mass, i.e.\ there is a mass gap.  In type II the critical solution is self-similar or scale-invariant and black holes can form with infinitesimally small masses, i.e.\ there is no mass gap.  For reviews see \cite{Gundlach:2002sx, Gundlach:2007gc}.  Type I and type II critical behavior has been extensively studied in spherical symmetry and found in numerous systems \cite{Gundlach:2002sx, Gundlach:2007gc}.  They were investigated in the related model of pure $SU(2)$ (no scalar field) in \cite{Choptuik:1996yg, Gundlach:1996je}.  A non-dynamical study of the monopole system related to type II behavior is given in \cite{Lue:1999zp, Lue:2000qr, Brihaye:1999kt}.    All indications are that the circles in Figs.\ \ref{fig:end state} (a), (d), and (g) represent type II collapse, but a proper identification of this requires more sophisticated numerical techniques than I am using here (such as adaptive mesh techniques). 

Less studied is another type of critical behavior found by Choptuik et.\ al.\ in \cite{Choptuik:1999gh} in which the critical solution sits between two types of black holes (as opposed to between collapse and non-collapse).  Analogously to types I and II, given one-parameter families of initial data from which two different spacetimes containing black holes are possible, for $p > p^*$ the end state of the evolution is one of the black hole spacetimes, while for $p < p^*$ the end state is the other black hole spacetime, and the critical solution at $p = p^*$ is an unstable attractor with a single decay mode.  Choptuik, Hirschmann, and Marsa \cite{Choptuik:1999gh} and Rinne \cite{Rinne} studied this phenomenon in pure $SU(2)$.  The two black hole spacetimes both contained Schwarzschild black holes but with different configurations for the gauge field.  In some parts of the initial data parameter space the critical solutions are the (fundamental) static black hole solutions found in \cite{Volkov:1989fi, Bizon:1990sr, Kunzle:1990is}, which have been shown to have a single decay mode with respect to radial perturbations \cite{Straumann:1990as}.  In other parts of parameter space the RN solution approximates a critical solution (it is approximate because the RN solution has an infinite number of decay modes but one of the modes can be tuned to dominate over the others).  Millward and Hirschmann \cite{Millward:2002pk} studied $SU(2)$ with a scalar field in the fundamental representation, i.e.\ as a complex doublet.  One of the black hole spacetimes was the Schwarzschild solution and the other was a sphaleron configuration with a black hole inside.  They too found critical solutions but did not compare them to known static solutions.

The monopole system is $SU(2)$ with a scalar field in the adjoint representation, i.e.\ as a real triplet.  One black hole spacetime is the RN solution and the other is a monopole configuration with a black hole inside, analogous to \cite{Millward:2002pk}.  Analogous to \cite{Choptuik:1999gh, Rinne} there exist well-known and well-studied static solutions for comparison with any critical solution.  Given the lack of study of this type of critical behavior (compared to the extensive study of type I and type II critical phenomena) I focus on the critical behavior between RN and monopole black holes.

\begin{figure*}
\centering
\includegraphics[width=6.5in]{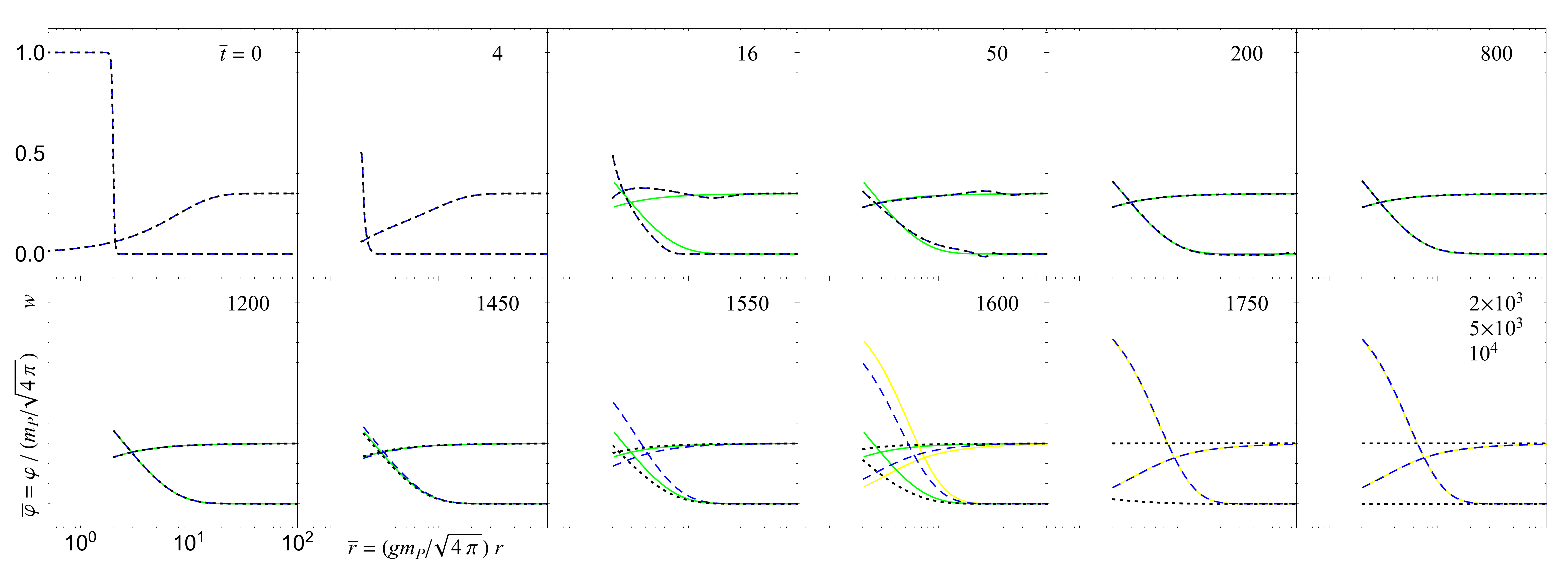}
\caption{Time evolution similar to Fig.\ \ref{fig:regular monopole}, but for two near-critical dynamic solutions with initial data using $(\bar{v}, \bar{\lambda}, \bar{x}_w, \bar{s}_\varphi) = (0.3, 0, 2.0, 10)$.  The $\bar{s}_w > \bar{s}_w^*$ solutions are the dashed blue curves and the $\bar{s}_w < \bar{s}_w^*$ are the dotted black curves, which begin on top of each other since for both $|\bar{s}_w - \bar{s}_w^*|/\bar{s}_w^* \approx 10^{-15}$.  The solid green lines introduced in the $\bar{t} =16$ frame are a static monopole black hole solution from the lower branch in Fig.\ \ref{figure2}(c) with $(\overline{M},\bar{r}_h) = (1.280, 2.077)$ and is an intermediate attractor since both dynamic solutions evolve toward it.  Eventually the dynamic solutions leave the green lines and separate as seen starting in the $\bar{t} = 1450$ frame.  The yellow lines introduced in the $\bar{t} = 1600$ frame are a static monopole black hole solution from the upper branch in Fig.\ \ref{figure2}(c) with $(\overline{M},\bar{r}_h) = (1.285, 2.098)$, which one of the dynamic solutions evolves to as its stable end state.  The other dynamic solution evolves to the RN black hole (\ref{RN BH sol}) as its stable end state.  Dynamic solutions for three different times are shown in the final frame.}
\label{fig:critical example}
\end{figure*}

In Fig.\ \ref{fig:critical example} I show a time evolution for two near-critical solutions.  Focusing for a moment on the $\bar{t}=0$ frame the $\bar{s}_w > \bar{s}_w^*$ solutions are plotted as the dashed blue curves and the $\bar{s}_w < \bar{s}_w^*$ solutions are plotted as the dotted black curves.  These solutions are seen to be directly on top of each other because they are both near-critical with $|\bar{s}_w - \bar{s}_w^*|/\bar{s}_w^* \approx 10^{-15}$.  As the evolution progresses they evolve together.  Starting in the $\bar{t} = 16$ frame I include as the solid green lines a static solution from the lower branch in Fig.\ \ref{figure2}(c).  The dynamic solutions are seen to evolve to it.  In frame $\bar{t} = 1450$ the dynamic solutions begin to move away from the static solution and from each other.  Starting in the $\bar{t} = 1600$ frame I include as the solid yellow lines a static solution from the upper branch in Fig.\ \ref{figure2}(c).  The dashed blue curves are seen to evolve to the solid yellow lines as their stable black hole monopole end state.  The dotted black curves are seen to evolve to the RN solution (\ref{RN BH sol}) as their stable end state.

The bottom monopole branches in Figs.\ \ref{figure2}(b-d) always act as attractors, with near critical solutions with $\bar{s}_w > \bar{s}_w^*$ decaying to monopole black holes on the upper branch and near critical solutions with $\bar{s}_w < \bar{s}_w^*$ decaying to RN black holes.  The values of $\bar{s}_w^*$ are indicated by the squares in Fig.\ \ref{fig:end state} and the evolution diagrams for all near-critical solutions are similar to Fig.\ \ref{fig:critical example}.  This corroborates the stability discussion in Sec.\ \ref{sec:static solutions} that the bottom branches in Figs.\ \ref{figure2}(b--d) are unstable, the top branches are stable, and the RN solutions for $\overline{M} > \overline{M}_B$, where $B$ is the bifurcation point, are stable.  This is also strong evidence that the lower branch monopole solutions are critical solutions.  (Identifying them as true critical solutions requires showing they have a single (radial) decay mode, which as far I am aware has not been done.)

\begin{figure*}
\centering
\includegraphics[width=6.5in]{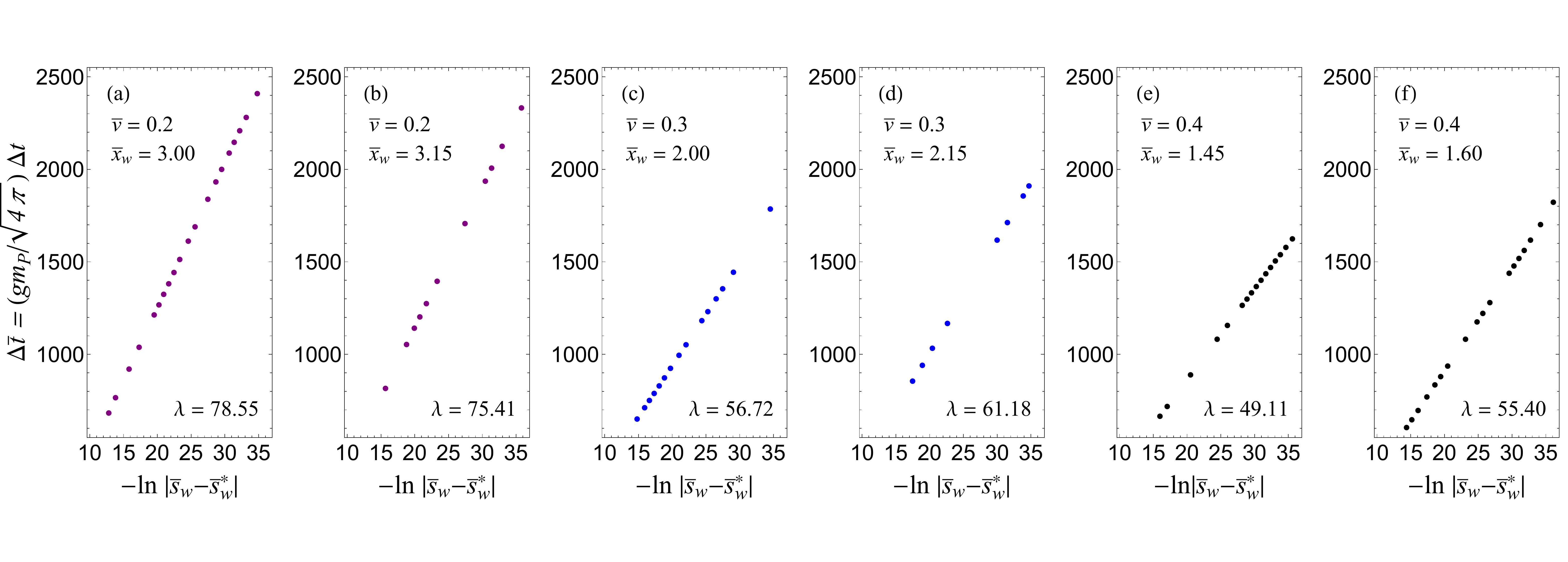}
\caption{Critical solutions sitting between stable black hole monopole and stable RN black hole end states obey the time scaling $\Delta \bar{t} = -\lambda \ln |\bar{s}_w - \bar{s}_w^*|$ where $\Delta \bar{t}$ is the elapsed time measured by an observer at infinity and $\lambda$ (not to be confused with the scalar field self-coupling) is the characteristic time scale for decay of the unstable critical solution or the inverse of the Lyapounov exponent for the unstable mode.  This time scaling expresses that the closer you get to the critical solution, i.e.\ the closer $\bar{s}_w$ gets to $\bar{s}_w^*$, the longer the dynamic solutions spend near the critical solution before evolving away.  I define $\Delta \bar{t}$ for the figures here to be the time from the beginning of the evolution until $|\bar{\varphi}(\bar{r}) - \bar{v}| < 10^{-3}$ for all unexcised values of $\bar{r}$ (I have found that once this inequality is satisfied an RN black hole end state is assured to occur).  Each figure is for the indicated values of $\bar{v}$ and $\bar{x}_w$.  Also shown in the figure is $\lambda$ as obtained from a least-squares fit.}
\label{fig:t v p}
\end{figure*}

When this type of critical behavior was discovered in \cite{Choptuik:1999gh} it was shown that it exhibits time scaling qualitatively similar to type I in that the closer the initial data are to that for the critical solution, i.e.\ the closer $p$ is to $p^*$, the longer the near-critical solution stays next to the critical solution as measured, say, by an observer at infinity, before decaying to its end state.  In terms of Fig.\ \ref{fig:critical example} this means the dynamic solutions spend more and more time on the solid green lines, as they do in frames $\bar{t} = 200$ to 1200,  as $\bar{s}_w$ approaches $\bar{s}_w^*$.  It was also shown in \cite{Choptuik:1999gh} that this time scaling obeys
\begin{equation}
\Delta \bar{t} = -\lambda \ln |p - p^*|
\end{equation}
where $\lambda$ (not to be confused with the scalar field self-coupling) is the characteristic time scale for decay of the unstable critical solution or the inverse of the Lyapounov exponent for the unstable mode.  This scaling relation was also found in \cite{Millward:2002pk, Rinne}.  In Fig.\ \ref{fig:t v p} I confirm this scaling relation for the monopole system and compute $\lambda$ for a number of critical solutions.


\subsection{No-Hair Conjecture}
\label{sec:hair}

The no-hair conjecture states that within a given model stationary black holes are uniquely determined by global charges that may be measured at infinity through surface integrals \cite{Ruffini:1971bza, Volkov:1998cc, Volkov:2016ehx}.  The solutions reviewed in Sec.\ \ref{sec:static solutions} are static (there is no angular momentum) and have unit magnetic charge and thus the only global parameter by which they could differ at infinity is mass.  That we have two stable solutions with the same mass (the upper branch monopole solution and the RN solution in the region $\overline{M}_B < \overline{M} < \overline{M}_A$ in Figs.\ \ref{figure2}(b--c)) is a well-known counter example to the no-hair conjecture \cite{Lee:1991vy, Breitenlohner:1991aa, Aichelburg:1992st} (see also \cite{Bizon:1992gb}).  

It is sometimes thought that the no-hair conjecture may still hold for black holes formed from collapse \cite{Volkov:1998cc}.  As we have seen in this section initial data can evolve to a stable end state that is either a black hole monopole or an RN black hole suggesting a dynamic counter-example to the no-hair conjecture.


\section{Conclusion}
\label{sec:conclusion}

I dynamically evolved spherically symmetric spacetimes containing gravitational 't Hooft--Polyakov monopoles.  Using black hole excision methods I determined the stable end states of the evolutions.  To reduce the large amount of parameter space that could be studied I focused on $\bar{\lambda} = 0$ and $\bar{v} = 0.2$, 0.3, and 0.4.  I also worked almost exclusively within the magnetic ansatz.  In this final section I comment on expectations for other ranges of parameters.  

The results for nonzero values of $\bar{\lambda}$ that are not too large I expect to be qualitatively similar, as the stability structure of the static solutions is the same \cite{Aichelburg:1992st, Tachizawa:1994wn}.  For larger $\lambda$ the unstable branch of the static black hole monopole solutions disappears and the RN solution, when it has the same mass as a monopole solution, is unstable.  This suggests that the critical behavior studied in Sec.\ \ref{sec:critical} disappears and the end state of an evolution in which a black hole forms will be a monopole black hole for masses in which it exists otherwise it will be an RN black hole.

As $\bar{v}$ increases eventually monopole and RN black holes no longer coexist (for example, when $\bar{v} > \sqrt{3}/2$ when $\bar{\lambda} = 0$ \cite{Breitenlohner:1991aa}).  There also exists a maximum value of $\bar{v}$ above which static monopole solutions do not exist \cite{Lee:1991vy, Breitenlohner:1991aa}.  In these regions we should not expect critical behavior as studied in Sec.\ \ref{sec:critical} and the stable end state should be whichever unique static black hole is possible.  In the opposite direction, for $\bar{v} \rightarrow 0$ the scalar field decouples and we have a pure $SU(2)$ system as studied in \cite{Choptuik:1996yg, Gundlach:1996je, Choptuik:1999gh, Rinne}.   Excited static monopole black holes (i.e.\ solutions with a nonzero number of nodes or zero-crossings of the gauge field, and which exist, for example, for $\bar{v} <\sqrt{3}/2$ when $\bar{\lambda} = 0$ \cite{Breitenlohner:1991aa}) are all expected to be unstable and it is unlikely they can be produced in a dynamic evolution.

Finally, everything done here is within the magnetic ansatz, as was the case in \cite{Choptuik:1999gh, Rinne, Millward:2002pk}.  At least for pure $SU(2)$ (no scalar field) this is known to be unstable to small (sphaleron) perturbations \cite{Volkov:1994dq, Lavrelashvili:1994rp, Volkov:1995np}, but since all data evolved here remain within the magnetic ansatz any instabilities in the sphaleron sector are unexcited.  It would be interesting to study this system without making the magnetic ansatz, as was done for pure $SU(2)$ in \cite{Rinne:2013qc}.



\begin{thebibliography}{99}%
\makeatletter
\providecommand \@ifxundefined [1]{%
 \@ifx{#1\undefined}
}%
\providecommand \@ifnum [1]{%
 \ifnum #1\expandafter \@firstoftwo
 \else \expandafter \@secondoftwo
 \fi
}%
\providecommand \@ifx [1]{%
 \ifx #1\expandafter \@firstoftwo
 \else \expandafter \@secondoftwo
 \fi
}%
\providecommand \natexlab [1]{#1}%
\providecommand \enquote  [1]{``#1''}%
\providecommand \bibnamefont  [1]{#1}%
\providecommand \bibfnamefont [1]{#1}%
\providecommand \citenamefont [1]{#1}%
\providecommand \href@noop [0]{\@secondoftwo}%
\providecommand \href [0]{\begingroup \@sanitize@url \@href}%
\providecommand \@href[1]{\@@startlink{#1}\@@href}%
\providecommand \@@href[1]{\endgroup#1\@@endlink}%
\providecommand \@sanitize@url [0]{\catcode `\\12\catcode `\$12\catcode
  `\&12\catcode `\#12\catcode `\^12\catcode `\_12\catcode `\%12\relax}%
\providecommand \@@startlink[1]{}%
\providecommand \@@endlink[0]{}%
\providecommand \url  [0]{\begingroup\@sanitize@url \@url }%
\providecommand \@url [1]{\endgroup\@href {#1}{\urlprefix }}%
\providecommand \urlprefix  [0]{URL }%
\providecommand \Eprint [0]{\href }%
\providecommand \doibase [0]{http://dx.doi.org/}%
\providecommand \selectlanguage [0]{\@gobble}%
\providecommand \bibinfo  [0]{\@secondoftwo}%
\providecommand \bibfield  [0]{\@secondoftwo}%
\providecommand \translation [1]{[#1]}%
\providecommand \BibitemOpen [0]{}%
\providecommand \bibitemStop [0]{}%
\providecommand \bibitemNoStop [0]{.\EOS\space}%
\providecommand \EOS [0]{\spacefactor3000\relax}%
\providecommand \BibitemShut  [1]{\csname bibitem#1\endcsname}%
\let\auto@bib@innerbib\@empty
\bibitem [{\citenamefont {Goddard}\ and\ \citenamefont
  {Olive}(1978)}]{Goddard:1977da}%
  \BibitemOpen
  \bibfield  {author} {\bibinfo {author} {\bibfnamefont {P.}~\bibnamefont
  {Goddard}}\ and\ \bibinfo {author} {\bibfnamefont {D.~I.}\ \bibnamefont
  {Olive}},\ }\href {\doibase 10.1088/0034-4885/41/9/001} {\bibfield  {journal}
  {\bibinfo  {journal} {Rept. Prog. Phys.}\ }\textbf {\bibinfo {volume} {41}},\
  \bibinfo {pages} {1357} (\bibinfo {year} {1978})}\BibitemShut {NoStop}%
\bibitem [{\citenamefont {'t~Hooft}(1974)}]{tHooft:1974kcl}%
  \BibitemOpen
  \bibfield  {author} {\bibinfo {author} {\bibfnamefont {G.}~\bibnamefont
  {'t~Hooft}},\ }\href {\doibase 10.1016/0550-3213(74)90486-6} {\bibfield
  {journal} {\bibinfo  {journal} {Nucl. Phys.}\ }\textbf {\bibinfo {volume}
  {B79}},\ \bibinfo {pages} {276} (\bibinfo {year} {1974})}\BibitemShut
  {NoStop}%
\bibitem [{\citenamefont {Polyakov}(1974)}]{Polyakov:1974ek}%
  \BibitemOpen
  \bibfield  {author} {\bibinfo {author} {\bibfnamefont {A.~M.}\ \bibnamefont
  {Polyakov}},\ }\href@noop {} {\bibfield  {journal} {\bibinfo  {journal} {JETP
  Lett.}\ }\textbf {\bibinfo {volume} {20}},\ \bibinfo {pages} {194} (\bibinfo
  {year} {1974})},\ \bibinfo {note} {[Pisma Zh. Eksp. Teor.
  Fiz.20,430(1974)]}\BibitemShut {NoStop}%
\bibitem [{\citenamefont {Van~Nieuwenhuizen}\ \emph {et~al.}(1976)\citenamefont
  {Van~Nieuwenhuizen}, \citenamefont {Wilkinson},\ and\ \citenamefont
  {Perry}}]{VanNieuwenhuizen:1975tc}%
  \BibitemOpen
  \bibfield  {author} {\bibinfo {author} {\bibfnamefont {P.}~\bibnamefont
  {Van~Nieuwenhuizen}}, \bibinfo {author} {\bibfnamefont {D.}~\bibnamefont
  {Wilkinson}}, \ and\ \bibinfo {author} {\bibfnamefont {M.~J.}\ \bibnamefont
  {Perry}},\ }\href {\doibase 10.1103/PhysRevD.13.778} {\bibfield  {journal}
  {\bibinfo  {journal} {Phys. Rev.}\ }\textbf {\bibinfo {volume} {D13}},\
  \bibinfo {pages} {778} (\bibinfo {year} {1976})}\BibitemShut {NoStop}%
\bibitem [{\citenamefont {Lee}\ \emph {et~al.}(1992{\natexlab{a}})\citenamefont
  {Lee}, \citenamefont {Nair},\ and\ \citenamefont {Weinberg}}]{Lee:1991vy}%
  \BibitemOpen
  \bibfield  {author} {\bibinfo {author} {\bibfnamefont {K.-M.}\ \bibnamefont
  {Lee}}, \bibinfo {author} {\bibfnamefont {V.~P.}\ \bibnamefont {Nair}}, \
  and\ \bibinfo {author} {\bibfnamefont {E.~J.}\ \bibnamefont {Weinberg}},\
  }\href {\doibase 10.1103/PhysRevD.45.2751} {\bibfield  {journal} {\bibinfo
  {journal} {Phys. Rev.}\ }\textbf {\bibinfo {volume} {D45}},\ \bibinfo {pages}
  {2751} (\bibinfo {year} {1992}{\natexlab{a}})},\ \Eprint
  {http://arxiv.org/abs/hep-th/9112008} {arXiv:hep-th/9112008 [hep-th]}
  \BibitemShut {NoStop}%
\bibitem [{\citenamefont {Ortiz}(1992)}]{Ortiz:1991eu}%
  \BibitemOpen
  \bibfield  {author} {\bibinfo {author} {\bibfnamefont {M.~E.}\ \bibnamefont
  {Ortiz}},\ }\href {\doibase 10.1103/PhysRevD.45.R2586} {\bibfield  {journal}
  {\bibinfo  {journal} {Phys. Rev.}\ }\textbf {\bibinfo {volume} {D45}},\
  \bibinfo {pages} {R2586} (\bibinfo {year} {1992})}\BibitemShut {NoStop}%
\bibitem [{\citenamefont {Breitenlohner}\ \emph {et~al.}(1992)\citenamefont
  {Breitenlohner}, \citenamefont {Forgacs},\ and\ \citenamefont
  {Maison}}]{Breitenlohner:1991aa}%
  \BibitemOpen
  \bibfield  {author} {\bibinfo {author} {\bibfnamefont {P.}~\bibnamefont
  {Breitenlohner}}, \bibinfo {author} {\bibfnamefont {P.}~\bibnamefont
  {Forgacs}}, \ and\ \bibinfo {author} {\bibfnamefont {D.}~\bibnamefont
  {Maison}},\ }\href {\doibase 10.1016/0550-3213(92)90682-2} {\bibfield
  {journal} {\bibinfo  {journal} {Nucl. Phys.}\ }\textbf {\bibinfo {volume}
  {B383}},\ \bibinfo {pages} {357} (\bibinfo {year} {1992})}\BibitemShut
  {NoStop}%
\bibitem [{\citenamefont {Breitenlohner}\ \emph {et~al.}(1995)\citenamefont
  {Breitenlohner}, \citenamefont {Forgacs},\ and\ \citenamefont
  {Maison}}]{Breitenlohner:1994di}%
  \BibitemOpen
  \bibfield  {author} {\bibinfo {author} {\bibfnamefont {P.}~\bibnamefont
  {Breitenlohner}}, \bibinfo {author} {\bibfnamefont {P.}~\bibnamefont
  {Forgacs}}, \ and\ \bibinfo {author} {\bibfnamefont {D.}~\bibnamefont
  {Maison}},\ }\href {\doibase 10.1016/S0550-3213(95)00100-X} {\bibfield
  {journal} {\bibinfo  {journal} {Nucl. Phys.}\ }\textbf {\bibinfo {volume}
  {B442}},\ \bibinfo {pages} {126} (\bibinfo {year} {1995})},\ \Eprint
  {http://arxiv.org/abs/gr-qc/9412039} {arXiv:gr-qc/9412039 [gr-qc]}
  \BibitemShut {NoStop}%
\bibitem [{\citenamefont {Volkov}\ and\ \citenamefont
  {Gal'tsov}(1999)}]{Volkov:1998cc}%
  \BibitemOpen
  \bibfield  {author} {\bibinfo {author} {\bibfnamefont {M.~S.}\ \bibnamefont
  {Volkov}}\ and\ \bibinfo {author} {\bibfnamefont {D.~V.}\ \bibnamefont
  {Gal'tsov}},\ }\href {\doibase 10.1016/S0370-1573(99)00010-1} {\bibfield
  {journal} {\bibinfo  {journal} {Phys. Rept.}\ }\textbf {\bibinfo {volume}
  {319}},\ \bibinfo {pages} {1} (\bibinfo {year} {1999})},\ \Eprint
  {http://arxiv.org/abs/hep-th/9810070} {arXiv:hep-th/9810070 [hep-th]}
  \BibitemShut {NoStop}%
\bibitem [{\citenamefont {Lee}\ \emph {et~al.}(1992{\natexlab{b}})\citenamefont
  {Lee}, \citenamefont {Nair},\ and\ \citenamefont {Weinberg}}]{Lee:1991qs}%
  \BibitemOpen
  \bibfield  {author} {\bibinfo {author} {\bibfnamefont {K.-M.}\ \bibnamefont
  {Lee}}, \bibinfo {author} {\bibfnamefont {V.~P.}\ \bibnamefont {Nair}}, \
  and\ \bibinfo {author} {\bibfnamefont {E.~J.}\ \bibnamefont {Weinberg}},\
  }\href {\doibase 10.1103/PhysRevLett.68.1100} {\bibfield  {journal} {\bibinfo
   {journal} {Phys. Rev. Lett.}\ }\textbf {\bibinfo {volume} {68}},\ \bibinfo
  {pages} {1100} (\bibinfo {year} {1992}{\natexlab{b}})},\ \Eprint
  {http://arxiv.org/abs/hep-th/9111045} {arXiv:hep-th/9111045 [hep-th]}
  \BibitemShut {NoStop}%
\bibitem [{\citenamefont {Aichelburg}\ and\ \citenamefont
  {Bizon}(1993)}]{Aichelburg:1992st}%
  \BibitemOpen
  \bibfield  {author} {\bibinfo {author} {\bibfnamefont {P.~C.}\ \bibnamefont
  {Aichelburg}}\ and\ \bibinfo {author} {\bibfnamefont {P.}~\bibnamefont
  {Bizon}},\ }\href {\doibase 10.1103/PhysRevD.48.607} {\bibfield  {journal}
  {\bibinfo  {journal} {Phys. Rev.}\ }\textbf {\bibinfo {volume} {D48}},\
  \bibinfo {pages} {607} (\bibinfo {year} {1993})},\ \Eprint
  {http://arxiv.org/abs/gr-qc/9212009} {arXiv:gr-qc/9212009 [gr-qc]}
  \BibitemShut {NoStop}%
\bibitem [{\citenamefont {Maeda}\ \emph {et~al.}(1994)\citenamefont {Maeda},
  \citenamefont {Tachizawa}, \citenamefont {Torii},\ and\ \citenamefont
  {Maki}}]{Maeda:1993ap}%
  \BibitemOpen
  \bibfield  {author} {\bibinfo {author} {\bibfnamefont {K.-I.}\ \bibnamefont
  {Maeda}}, \bibinfo {author} {\bibfnamefont {T.}~\bibnamefont {Tachizawa}},
  \bibinfo {author} {\bibfnamefont {T.}~\bibnamefont {Torii}}, \ and\ \bibinfo
  {author} {\bibfnamefont {T.}~\bibnamefont {Maki}},\ }\href {\doibase
  10.1103/PhysRevLett.72.450} {\bibfield  {journal} {\bibinfo  {journal} {Phys.
  Rev. Lett.}\ }\textbf {\bibinfo {volume} {72}},\ \bibinfo {pages} {450}
  (\bibinfo {year} {1994})},\ \Eprint {http://arxiv.org/abs/gr-qc/9310015}
  {arXiv:gr-qc/9310015 [gr-qc]} \BibitemShut {NoStop}%
\bibitem [{\citenamefont {Tachizawa}\ \emph {et~al.}(1995)\citenamefont
  {Tachizawa}, \citenamefont {Maeda},\ and\ \citenamefont
  {Torii}}]{Tachizawa:1994wn}%
  \BibitemOpen
  \bibfield  {author} {\bibinfo {author} {\bibfnamefont {T.}~\bibnamefont
  {Tachizawa}}, \bibinfo {author} {\bibfnamefont {K.-I.}\ \bibnamefont
  {Maeda}}, \ and\ \bibinfo {author} {\bibfnamefont {T.}~\bibnamefont
  {Torii}},\ }\href {\doibase 10.1103/PhysRevD.51.4054} {\bibfield  {journal}
  {\bibinfo  {journal} {Phys. Rev.}\ }\textbf {\bibinfo {volume} {D51}},\
  \bibinfo {pages} {4054} (\bibinfo {year} {1995})},\ \Eprint
  {http://arxiv.org/abs/gr-qc/9410016} {arXiv:gr-qc/9410016 [gr-qc]}
  \BibitemShut {NoStop}%
\bibitem [{\citenamefont {Hollmann}(1994)}]{Hollmann:1994fm}%
  \BibitemOpen
  \bibfield  {author} {\bibinfo {author} {\bibfnamefont {H.}~\bibnamefont
  {Hollmann}},\ }\href {\doibase 10.1016/0370-2693(94)91364-1} {\bibfield
  {journal} {\bibinfo  {journal} {Phys. Lett.}\ }\textbf {\bibinfo {volume}
  {B338}},\ \bibinfo {pages} {181} (\bibinfo {year} {1994})},\ \Eprint
  {http://arxiv.org/abs/gr-qc/9406018} {arXiv:gr-qc/9406018 [gr-qc]}
  \BibitemShut {NoStop}%
\bibitem [{\citenamefont {Choptuik}(1993)}]{Choptuik:1992jv}%
  \BibitemOpen
  \bibfield  {author} {\bibinfo {author} {\bibfnamefont {M.~W.}\ \bibnamefont
  {Choptuik}},\ }\href {\doibase 10.1103/PhysRevLett.70.9} {\bibfield
  {journal} {\bibinfo  {journal} {Phys. Rev. Lett.}\ }\textbf {\bibinfo
  {volume} {70}},\ \bibinfo {pages} {9} (\bibinfo {year} {1993})}\BibitemShut
  {NoStop}%
\bibitem [{\citenamefont {Choptuik}\ \emph {et~al.}(1996)\citenamefont
  {Choptuik}, \citenamefont {Chmaj},\ and\ \citenamefont
  {Bizon}}]{Choptuik:1996yg}%
  \BibitemOpen
  \bibfield  {author} {\bibinfo {author} {\bibfnamefont {M.~W.}\ \bibnamefont
  {Choptuik}}, \bibinfo {author} {\bibfnamefont {T.}~\bibnamefont {Chmaj}}, \
  and\ \bibinfo {author} {\bibfnamefont {P.}~\bibnamefont {Bizon}},\ }\href
  {\doibase 10.1103/PhysRevLett.77.424} {\bibfield  {journal} {\bibinfo
  {journal} {Phys. Rev. Lett.}\ }\textbf {\bibinfo {volume} {77}},\ \bibinfo
  {pages} {424} (\bibinfo {year} {1996})},\ \Eprint
  {http://arxiv.org/abs/gr-qc/9603051} {arXiv:gr-qc/9603051 [gr-qc]}
  \BibitemShut {NoStop}%
\bibitem [{\citenamefont {Gundlach}(2003)}]{Gundlach:2002sx}%
  \BibitemOpen
  \bibfield  {author} {\bibinfo {author} {\bibfnamefont {C.}~\bibnamefont
  {Gundlach}},\ }\href {\doibase 10.1016/S0370-1573(02)00560-4} {\bibfield
  {journal} {\bibinfo  {journal} {Phys. Rept.}\ }\textbf {\bibinfo {volume}
  {376}},\ \bibinfo {pages} {339} (\bibinfo {year} {2003})},\ \Eprint
  {http://arxiv.org/abs/gr-qc/0210101} {arXiv:gr-qc/0210101 [gr-qc]}
  \BibitemShut {NoStop}%
\bibitem [{\citenamefont {Gundlach}\ and\ \citenamefont
  {Martin-Garcia}(2007)}]{Gundlach:2007gc}%
  \BibitemOpen
  \bibfield  {author} {\bibinfo {author} {\bibfnamefont {C.}~\bibnamefont
  {Gundlach}}\ and\ \bibinfo {author} {\bibfnamefont {J.~M.}\ \bibnamefont
  {Martin-Garcia}},\ }\href {\doibase 10.12942/lrr-2007-5} {\bibfield
  {journal} {\bibinfo  {journal} {Living Rev. Rel.}\ }\textbf {\bibinfo
  {volume} {10}},\ \bibinfo {pages} {5} (\bibinfo {year} {2007})},\ \Eprint
  {http://arxiv.org/abs/0711.4620} {arXiv:0711.4620 [gr-qc]} \BibitemShut
  {NoStop}%
\bibitem [{\citenamefont {Choptuik}\ \emph {et~al.}(1999)\citenamefont
  {Choptuik}, \citenamefont {Hirschmann},\ and\ \citenamefont
  {Marsa}}]{Choptuik:1999gh}%
  \BibitemOpen
  \bibfield  {author} {\bibinfo {author} {\bibfnamefont {M.~W.}\ \bibnamefont
  {Choptuik}}, \bibinfo {author} {\bibfnamefont {E.~W.}\ \bibnamefont
  {Hirschmann}}, \ and\ \bibinfo {author} {\bibfnamefont {R.~L.}\ \bibnamefont
  {Marsa}},\ }\href {\doibase 10.1103/PhysRevD.60.124011} {\bibfield  {journal}
  {\bibinfo  {journal} {Phys. Rev.}\ }\textbf {\bibinfo {volume} {D60}},\
  \bibinfo {pages} {124011} (\bibinfo {year} {1999})},\ \Eprint
  {http://arxiv.org/abs/gr-qc/9903081} {arXiv:gr-qc/9903081 [gr-qc]}
  \BibitemShut {NoStop}%
\bibitem [{\citenamefont {Millward}\ and\ \citenamefont
  {Hirschmann}(2003)}]{Millward:2002pk}%
  \BibitemOpen
  \bibfield  {author} {\bibinfo {author} {\bibfnamefont {R.~S.}\ \bibnamefont
  {Millward}}\ and\ \bibinfo {author} {\bibfnamefont {E.~W.}\ \bibnamefont
  {Hirschmann}},\ }\href {\doibase 10.1103/PhysRevD.68.024017} {\bibfield
  {journal} {\bibinfo  {journal} {Phys. Rev.}\ }\textbf {\bibinfo {volume}
  {D68}},\ \bibinfo {pages} {024017} (\bibinfo {year} {2003})},\ \Eprint
  {http://arxiv.org/abs/gr-qc/0212015} {arXiv:gr-qc/0212015 [gr-qc]}
  \BibitemShut {NoStop}%
\bibitem{Rinne}
  O.~Rinne,
  Phys.\ Rev.\ D {\bf 90}, 124084 (2014),
  arXiv:1409.6173 [gr-qc].
\bibitem [{\citenamefont {Sakai}(1996)}]{Sakai:1995ds}%
  \BibitemOpen
  \bibfield  {author} {\bibinfo {author} {\bibfnamefont {N.}~\bibnamefont
  {Sakai}},\ }\href {\doibase 10.1103/PhysRevD.54.1548} {\bibfield  {journal}
  {\bibinfo  {journal} {Phys. Rev.}\ }\textbf {\bibinfo {volume} {D54}},\
  \bibinfo {pages} {1548} (\bibinfo {year} {1996})},\ \Eprint
  {http://arxiv.org/abs/gr-qc/9512045} {arXiv:gr-qc/9512045 [gr-qc]}
  \BibitemShut {NoStop}%
\bibitem{Fodor1} 
  G.~Fodor and I.~Racz,
  Phys.\ Rev.\ Lett.\  {\bf 92}, 151801 (2004), arXiv:hep-th/0311061.  
\bibitem{Fodor2} 
  G.~Fodor and I.~Racz,
  Phys.\ Rev.\ D {\bf 77}, 025019 (2008), arXiv:hep-th/0609110.
\bibitem{AlcubierreBook}
	M.\ Alcubierre, \textit{Introduction to 3+1 numerical relativity}, Oxford:\ Oxford University Press (2008).
\bibitem{BaumgarteBook}
	T.\ W.\ Baumgarte and S.\ L\ Shapiro, \textit{Numerical relativity:\ Solving Einstein's equations on the computer}, Cambridge:\ Cambridge University Press (2010).
\bibitem{Thornburg}
	J.\ Thornburg, Ph.D. thesis, University of British Columbia (1993).
\bibitem [{\citenamefont {Seidel}\ and\ \citenamefont
  {Suen}(1992)}]{Seidel:1992vd}%
  \BibitemOpen
  \bibfield  {author} {\bibinfo {author} {\bibfnamefont {E.}~\bibnamefont
  {Seidel}}\ and\ \bibinfo {author} {\bibfnamefont {W.-M.}\ \bibnamefont
  {Suen}},\ }\href {\doibase 10.1103/PhysRevLett.69.1845} {\bibfield  {journal}
  {\bibinfo  {journal} {Phys. Rev. Lett.}\ }\textbf {\bibinfo {volume} {69}},\
  \bibinfo {pages} {1845} (\bibinfo {year} {1992})},\ \Eprint
  {http://arxiv.org/abs/gr-qc/9210016} {arXiv:gr-qc/9210016 [gr-qc]}
  \BibitemShut {NoStop}%
\bibitem [{\citenamefont {Witten}(1977)}]{Witten:1976ck}%
  \BibitemOpen
  \bibfield  {author} {\bibinfo {author} {\bibfnamefont {E.}~\bibnamefont
  {Witten}},\ }\href {\doibase 10.1103/PhysRevLett.38.121} {\bibfield
  {journal} {\bibinfo  {journal} {Phys. Rev. Lett.}\ }\textbf {\bibinfo
  {volume} {38}},\ \bibinfo {pages} {121} (\bibinfo {year} {1977})}\BibitemShut
  {NoStop}%
\bibitem [{\citenamefont {Bartnik}\ and\ \citenamefont
  {McKinnon}(1988)}]{Bartnik:1988am}%
  \BibitemOpen
  \bibfield  {author} {\bibinfo {author} {\bibfnamefont {R.}~\bibnamefont
  {Bartnik}}\ and\ \bibinfo {author} {\bibfnamefont {J.}~\bibnamefont
  {McKinnon}},\ }\href {\doibase 10.1103/PhysRevLett.61.141} {\bibfield
  {journal} {\bibinfo  {journal} {Phys. Rev. Lett.}\ }\textbf {\bibinfo
  {volume} {61}},\ \bibinfo {pages} {141} (\bibinfo {year} {1988})}\BibitemShut
  {NoStop}%
\bibitem [{\citenamefont {Gundlach}(1997)}]{Gundlach:1996je}%
  \BibitemOpen
  \bibfield  {author} {\bibinfo {author} {\bibfnamefont {C.}~\bibnamefont
  {Gundlach}},\ }\href {\doibase 10.1103/PhysRevD.55.6002} {\bibfield
  {journal} {\bibinfo  {journal} {Phys. Rev.}\ }\textbf {\bibinfo {volume}
  {D55}},\ \bibinfo {pages} {6002} (\bibinfo {year} {1997})},\ \Eprint
  {http://arxiv.org/abs/gr-qc/9610069} {arXiv:gr-qc/9610069 [gr-qc]}
  \BibitemShut {NoStop}%
\bibitem [{\citenamefont {Lue}\ and\ \citenamefont
  {Weinberg}(1999)}]{Lue:1999zp}%
  \BibitemOpen
  \bibfield  {author} {\bibinfo {author} {\bibfnamefont {A.}~\bibnamefont
  {Lue}}\ and\ \bibinfo {author} {\bibfnamefont {E.~J.}\ \bibnamefont
  {Weinberg}},\ }\href {\doibase 10.1103/PhysRevD.60.084025} {\bibfield
  {journal} {\bibinfo  {journal} {Phys. Rev.}\ }\textbf {\bibinfo {volume}
  {D60}},\ \bibinfo {pages} {084025} (\bibinfo {year} {1999})},\ \Eprint
  {http://arxiv.org/abs/hep-th/9905223} {arXiv:hep-th/9905223 [hep-th]}
  \BibitemShut {NoStop}%
\bibitem [{\citenamefont {Lue}\ and\ \citenamefont
  {Weinberg}(2000)}]{Lue:2000qr}%
  \BibitemOpen
  \bibfield  {author} {\bibinfo {author} {\bibfnamefont {A.}~\bibnamefont
  {Lue}}\ and\ \bibinfo {author} {\bibfnamefont {E.~J.}\ \bibnamefont
  {Weinberg}},\ }\href {\doibase 10.1103/PhysRevD.61.124003} {\bibfield
  {journal} {\bibinfo  {journal} {Phys. Rev.}\ }\textbf {\bibinfo {volume}
  {D61}},\ \bibinfo {pages} {124003} (\bibinfo {year} {2000})},\ \Eprint
  {http://arxiv.org/abs/hep-th/0001140} {arXiv:hep-th/0001140 [hep-th]}
  \BibitemShut {NoStop}%
\bibitem [{\citenamefont {Brihaye}\ \emph {et~al.}(2000)\citenamefont
  {Brihaye}, \citenamefont {Hartmann},\ and\ \citenamefont
  {Kunz}}]{Brihaye:1999kt}%
  \BibitemOpen
  \bibfield  {author} {\bibinfo {author} {\bibfnamefont {Y.}~\bibnamefont
  {Brihaye}}, \bibinfo {author} {\bibfnamefont {B.}~\bibnamefont {Hartmann}}, \
  and\ \bibinfo {author} {\bibfnamefont {J.}~\bibnamefont {Kunz}},\ }\href
  {\doibase 10.1103/PhysRevD.62.044008} {\bibfield  {journal} {\bibinfo
  {journal} {Phys. Rev.}\ }\textbf {\bibinfo {volume} {D62}},\ \bibinfo {pages}
  {044008} (\bibinfo {year} {2000})},\ \Eprint
  {http://arxiv.org/abs/hep-th/9911148} {arXiv:hep-th/9911148 [hep-th]}
  \BibitemShut {NoStop}%
\bibitem [{\citenamefont {Volkov}\ and\ \citenamefont
  {Galtsov}(1989)}]{Volkov:1989fi}%
  \BibitemOpen
  \bibfield  {author} {\bibinfo {author} {\bibfnamefont {M.~S.}\ \bibnamefont
  {Volkov}}\ and\ \bibinfo {author} {\bibfnamefont {D.~V.}\ \bibnamefont
  {Galtsov}},\ }\href@noop {} {\bibfield  {journal} {\bibinfo  {journal} {JETP
  Lett.}\ }\textbf {\bibinfo {volume} {50}},\ \bibinfo {pages} {346} (\bibinfo
  {year} {1989})},\ \bibinfo {note} {[Pisma Zh. Eksp. Teor.
  Fiz.50,312(1989)]}\BibitemShut {NoStop}%
\bibitem [{\citenamefont {Bizon}(1990)}]{Bizon:1990sr}%
  \BibitemOpen
  \bibfield  {author} {\bibinfo {author} {\bibfnamefont {P.}~\bibnamefont
  {Bizon}},\ }\href {\doibase 10.1103/PhysRevLett.64.2844} {\bibfield
  {journal} {\bibinfo  {journal} {Phys. Rev. Lett.}\ }\textbf {\bibinfo
  {volume} {64}},\ \bibinfo {pages} {2844} (\bibinfo {year}
  {1990})}\BibitemShut {NoStop}%
\bibitem [{\citenamefont {K{\"u}nzle}\ and\ \citenamefont {Masood-ul
  Alam}(1990)}]{Kunzle:1990is}%
  \BibitemOpen
  \bibfield  {author} {\bibinfo {author} {\bibfnamefont {H.~P.}\ \bibnamefont
  {K{\"u}nzle}}\ and\ \bibinfo {author} {\bibfnamefont {A.~K.~M.}\ \bibnamefont
  {Masood-ul Alam}},\ }\href {\doibase 10.1063/1.528773} {\bibfield  {journal}
  {\bibinfo  {journal} {J. Math. Phys.}\ }\textbf {\bibinfo {volume} {31}},\
  \bibinfo {pages} {928} (\bibinfo {year} {1990})}\BibitemShut {NoStop}%
\bibitem [{\citenamefont {Straumann}\ and\ \citenamefont
  {Zhou}(1990)}]{Straumann:1990as}%
  \BibitemOpen
  \bibfield  {author} {\bibinfo {author} {\bibfnamefont {N.}~\bibnamefont
  {Straumann}}\ and\ \bibinfo {author} {\bibfnamefont {Z.~H.}\ \bibnamefont
  {Zhou}},\ }\href {\doibase 10.1016/0370-2693(90)90951-2} {\bibfield
  {journal} {\bibinfo  {journal} {Phys. Lett.}\ }\textbf {\bibinfo {volume}
  {B243}},\ \bibinfo {pages} {33} (\bibinfo {year} {1990})}\BibitemShut
  {NoStop}%
\bibitem [{\citenamefont {Ruffini}\ and\ \citenamefont
  {Wheeler}(1971)}]{Ruffini:1971bza}%
  \BibitemOpen
  \bibfield  {author} {\bibinfo {author} {\bibfnamefont {R.}~\bibnamefont
  {Ruffini}}\ and\ \bibinfo {author} {\bibfnamefont {J.~A.}\ \bibnamefont
  {Wheeler}},\ }\href {\doibase 10.1063/1.3022513} {\bibfield  {journal}
  {\bibinfo  {journal} {Phys. Today}\ }\textbf {\bibinfo {volume} {24}},\
  \bibinfo {pages} {30} (\bibinfo {year} {1971})}\BibitemShut {NoStop}%
\bibitem [{\citenamefont {Volkov}(2016)}]{Volkov:2016ehx}%
  \BibitemOpen
  \bibfield  {author} {\bibinfo {author} {\bibfnamefont {M.~S.}\ \bibnamefont
  {Volkov}},\ }\href@noop {} {\  (\bibinfo {year} {2016})},\ \Eprint
  {http://arxiv.org/abs/1601.08230} {arXiv:1601.08230 [gr-qc]} \BibitemShut
  {NoStop}%
\bibitem [{\citenamefont {Volkov}\ and\ \citenamefont
  {Galtsov}(1995)}]{Volkov:1994dq}%
  \BibitemOpen
  \bibfield  {author} {\bibinfo {author} {\bibfnamefont {M.~S.}\ \bibnamefont
  {Volkov}}\ and\ \bibinfo {author} {\bibfnamefont {D.~V.}\ \bibnamefont
  {Galtsov}},\ }\href {\doibase 10.1016/0370-2693(94)01310-9,
  10.1016/0370-2693(95)80005-I} {\bibfield  {journal} {\bibinfo  {journal}
  {Phys. Lett.}\ }\textbf {\bibinfo {volume} {B341}},\ \bibinfo {pages} {279}
  (\bibinfo {year} {1995})},\ \Eprint {http://arxiv.org/abs/hep-th/9409041}
  {arXiv:hep-th/9409041 [hep-th]} \BibitemShut {NoStop}%
\bibitem [{\citenamefont {Lavrelashvili}\ and\ \citenamefont
  {Maison}(1995)}]{Lavrelashvili:1994rp}%
  \BibitemOpen
  \bibfield  {author} {\bibinfo {author} {\bibfnamefont {G.~V.}\ \bibnamefont
  {Lavrelashvili}}\ and\ \bibinfo {author} {\bibfnamefont {D.}~\bibnamefont
  {Maison}},\ }\href {\doibase 10.1016/0370-2693(94)01479-V} {\bibfield
  {journal} {\bibinfo  {journal} {Phys. Lett.}\ }\textbf {\bibinfo {volume}
  {B343}},\ \bibinfo {pages} {214} (\bibinfo {year} {1995})},\ \Eprint
  {http://arxiv.org/abs/hep-th/9409185} {arXiv:hep-th/9409185 [hep-th]}
  \BibitemShut {NoStop}%
\bibitem [{\citenamefont {Volkov}\ \emph {et~al.}(1995)\citenamefont {Volkov},
  \citenamefont {Brodbeck}, \citenamefont {Lavrelashvili},\ and\ \citenamefont
  {Straumann}}]{Volkov:1995np}%
  \BibitemOpen
  \bibfield  {author} {\bibinfo {author} {\bibfnamefont {M.~S.}\ \bibnamefont
  {Volkov}}, \bibinfo {author} {\bibfnamefont {O.}~\bibnamefont {Brodbeck}},
  \bibinfo {author} {\bibfnamefont {G.~V.}\ \bibnamefont {Lavrelashvili}}, \
  and\ \bibinfo {author} {\bibfnamefont {N.}~\bibnamefont {Straumann}},\ }\href
  {\doibase 10.1016/0370-2693(95)00293-T} {\bibfield  {journal} {\bibinfo
  {journal} {Phys. Lett.}\ }\textbf {\bibinfo {volume} {B349}},\ \bibinfo
  {pages} {438} (\bibinfo {year} {1995})},\ \Eprint
  {http://arxiv.org/abs/hep-th/9502045} {arXiv:hep-th/9502045 [hep-th]}
  \BibitemShut {NoStop}%
\bibitem{Bizon:1992gb} 
  P.~Bizon and T.~Chmaj,
  Phys.\ Lett.\ B {\bf 297}, 55 (1992).
\bibitem{Rinne:2013qc}
  O.~Rinne and V.~Moncrief,
  Class.\ Quant.\ Grav.\  {\bf 30} (2013),
  arXiv:1301.6174 [gr-qc].
\end{thebibliography}

%

\end{document}